\documentclass[useAMS,usenatbib]{mn2e}

\usepackage{graphicx}
\usepackage{longtable}

\title[New brown dwarfs in Upper Sco from UKIDSS]{
New brown dwarfs in Upper Sco using UKIDSS Galactic Cluster Survey
science verification data
\thanks{Based on observations made with the United Kingdom Infrared 
Telescope, operated by the Joint Astronomy Centre on behalf of the 
U.K. Particle Physics and Astronomy Research Council.}}
\author[N. Lodieu et al.]{N. Lodieu$^{1}$\thanks{E-mail: nl41@star.le.ac.uk}
N. C. Hambly$^{2}$, R. F. Jameson$^{1}$, S. T. Hodgkin$^{3}$, G. Carraro$^{4,5}$ and
\newauthor 
T.R. Kendall$^{6}$  \\
$^{1}$Department of Physics \& Astronomy, University of Leicester, 
University Road, Leicester LE1 7RH, UK \\ 
$^{2}$Scottish Universities' Physics Alliance (SUPA), 
Institute for Astronomy, School of Physics, University of Edinburgh, \\
~Royal Observatory, Blackford Hill, 
Edinburgh EH9 3HJ, UK  \\
$^{3}$Institute of Astronomy, Madingley Road, Cambridge, CB3 0HA, UK \\
$^{4}$Departamento de Astronomía, Universidad de Chile, Casilla 36-D, 
Santiago, Chile \\
$^{5}$Astronomy Department, Yale University, PO Box 208101, 
New Haven, CT 06520-8101, USA \\ 
$^{6}$Centre for Astrophysics Research, Science \& Technology 
Research Institute, Department of Physics Astronomy \& Mathematics, \\
~University of Hertfordshire, College Lane, 
Hatfield, Hertfordshire AL10 9AB \\
}
\begin{document}

\date{Accepted ---. Received ---; in original form ---}

\pagerange{\pageref{firstpage}--\pageref{lastpage}} \pubyear{2005}

\maketitle

\label{firstpage}

%
%
\begin{abstract}
We present first results from a deep ($J$ = 18.7), wide-field (6.5 square 
degrees) infrared ($ZYJHK$) survey in the Upper Sco association 
conducted within the science verification phase of the UKIRT Infrared Deep
Sky Survey Galactic Cluster Survey (GCS).
Cluster members define a sequence well separated from field
stars in the ($Z-J$,$Z$) colour-magnitude diagram.
We have selected a total of 164 candidates with $J$ = 10.5--18.7 mag
from the ($Z-J$,$Z$) and ($Y-J$,$Y$) diagrams.
We further investigated the location of those candidates in the other
colour--magnitude and colour--colour diagrams to weed out contaminants.
The cross--correlation of the GCS catalogue with the 2MASS database 
confirms the membership of 116 photometric candidates down to 20 
Jupiter masses as they lie within a 2$\sigma$ circle centred on the
association mean motion. The final list of cluster members contains
129 sources with masses between 0.3 and 0.007 M$_{\odot}$.
We extracted a dozen new low-mass brown dwarfs below 
20 M$_{\rm Jup}$, the limit of previous surveys in the region.
Finally, we have derived the mass function in Upper Sco over the
0.3--0.01 M$_{\odot}$ mass range, best fit by a single segment 
with a slope of index $\alpha$ = 0.6$\pm$0.1, in agreement with 
previous determination in open clusters. 
\end{abstract}


%
%
\section{Introduction}
\label{USco$sv:intro}

The Scorpius Centaurus is the nearest OB association, located 
at a distance of 145$\pm$2 pc from the Sun \citep{deBruijne97,deZeeuw99}.
The association spans 120 square degrees and is composed 
of 3 subgroups: Upper Scorpius, 
Upper Centaurus Lupus, and Lower Centaurus Crux (Blaauw 1964). 
The age of the Upper Sco association is about 5 Myr with little scatter
\citep{preibisch02}. The region is free of extinction with 
Av $\leq$ 2 mag, suggesting that star formation has already ended 
\citep{walter94}.

The association has been studied at multiple wavelengths over the 
past decade. \citet{walter94} identified 28 low--mass stars over
7 deg$^{2}$ in Upper Sco from an X-ray survey performed
with the Einstein satellite. \citet{preibisch98} performed
a larger scale X-ray survey of 160 square degrees with ROSAT
and obtained spectroscopy for selected candidates, suggesting that
39 of them are indeed genuine members of the association.
\citet{deBruijne97} extracted 115 additional members from 
trigonometric parallaxes and kinematic information available 
for a sample of 1215 Hipparcos stars.
\citet{preibisch01} and \citet{preibisch02} published a sample
of spectroscopically confirmed X-ray selected members in the 
range 2.0~to~0.1 M$_{\odot}$ over 6 square degrees. The derived mass
function (not corrected for binaries) is in agreement with
the field Initial Mass Function \citep[IMF;][]{scalo98,kroupa02}.
The hydrogen burning boundary was crossed with the discovery of
64 very low--mass stars and brown dwarfs with spectral types
later than M5.5 by independent studies 
\citep{ardila00,mohanty04a,martin04,slesnick06}.

The UKIRT Infrared Deep Sky Survey 
\citep[UKIDSS;][]{lawrence06} 
constitutes a new generation of deep, large--scale infrared surveys. 
It consists of 5 components: the Large Area Survey, 
the Galactic Cluster Survey (hereafter GCS), 
the Galactic Plane Survey, the Deep Extragalactic Survey, 
and the Ultra-Deep Survey.
The GCS will cover $\sim1000$ square degrees in 10 star forming 
regions and open clusters down to $K$ = 18.4  at two epochs.
The main scientific driver of the survey is to study the 
Initial Mass Function and its dependence with environment
in the substellar regime using a uniformly selected data set of low--mass
stars and brown dwarfs. Ultimately,
the full coverage of the targets and the availability of
two epochs for proper motion measurement
will provide a complete census of very low--mass
stars, brown dwarfs, and planetary mass objects.

In this paper we analyse a 6.5 square degree survey conducted in the
Upper Sco association during the Science Verification phase 
of the UKIDSS project.
In Sect.\ \ref{USco:ukidss_GCS} we briefly present the observations,
the catalogue products, and the selection of sources.
In Sect.\ \ref{USco:cand} we describe the photometric selection
of cluster candidates in Upper Sco from various 
colour--magnitude 
diagrams and estimate proper motions for the brighter stars 
using 2MASS \citep{cutri03} to provide first epoch positions.
We also discuss possible contamination of our sample and
conclude that it will be very small.
In Sect.\ \ref{USco:discuss} we discuss the main results of the
survey, including the mass function down to
10 Jupiter masses.
Finally, we give our conclusions in Sect.\ \ref{USco:concl}.

%
%
\section{The UKIDSS GCS in Upper Sco}
\label{USco:ukidss_GCS}

\begin{figure}
   \includegraphics[width=\linewidth]{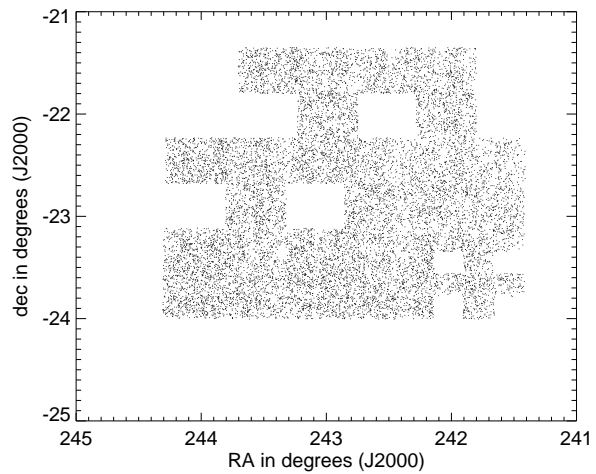}
   \caption{
            Tile coverage from the UKIDSS GCS Science Verification
            in the Upper Sco association. The total area covered
            is 6.5 square degrees. Note the somewhat uneven tile
            coverage from the GCS science verification phase.
}
   \label{fig_USco:coverage1}
\end{figure}

Eight tiles were observed with the UKIRT Wide--Field CAMera (WFCAM) 
in the association during the Science Verification phase on 
2005 April 8--13  
covering a total of 6.5 square degrees
(Fig.\ \ref{fig_USco:coverage1}) in $ZYJHK$ broadband filters
\citep{hewett06};
details of the filters are given in Table \ref{tab_USco:filters}.
The $Z$ WFCAM filter
is narrower than optical $z$ filters while the $Y$ filter
was specifically designed to uncover the coolest brown dwarfs
and furthest quasars, those being the
main scientific goals of the Large Area Survey.
In particular, a $Y-J \geq$ 1 selects L and T dwarfs.
The WFCAM focal plane array includes 4 Rockwell 2048\,$\times$\,2048 
chips each covering a 13.6 arcmin by 13.6 arcmin field 
resulting in a non-contiguous `pawprint' 
with a pixel scale of 0.4 arcsec. Each chip is spaced by $\sim95$\% of
the device size implying that four pawprints are required 
to obtain contiguous coverage in a tile of 0.8 square degrees
(Casali et al.\ 2006, in preparation). 

\begin{figure}
   \centering
   \includegraphics[width=\linewidth]{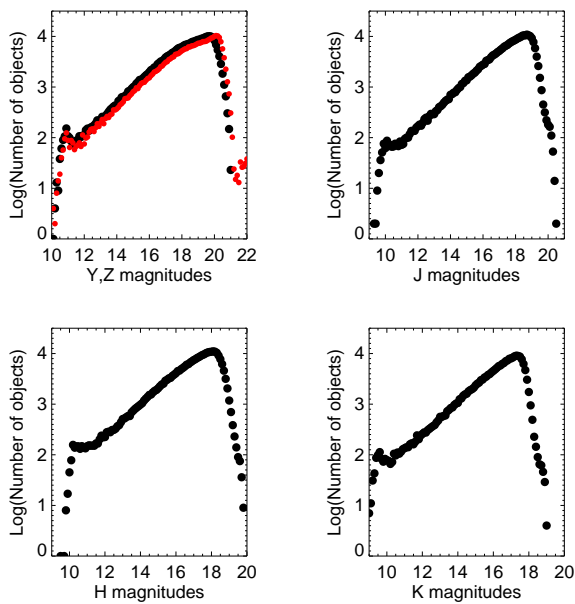}
   \caption{
Number--magnitude histograms for Upper Sco observations
in all filters,
including $Z$ and $Y$ (top left), $J$ (top right),
$H$ (bottom left), and $K$ (bottom right).
The 100\% completeness limits are estimated to $Z$ = 20.1, 
$Y$ = 19.8, $J$ = 18.7, $H$ = 18.1, $K$ = 17.3, respectively.
The completeness limits are defined as the magnitude where the
histogram stops following a straight line in a logarithmic scale.
}
   \label{fig_USco:completeness}
\end{figure}

A summary of the observations, including the 8 WFCAM pointings, central 
coordinates, date of observations, weather conditions, and seeing
is provided in Tables~\ref{tab_USco:log_obs} and~\ref{qc}.
Exposure times were set to 40 seconds in all passbands, 
yielding 100\% detection completeness limits of
$Z$ = 20.1, $Y$ = 19.8, $J$ = 18.7, $H$ = 18.1, $K$ = 17.3 mag
in the case of the science verification data in Upper Sco 
(Fig.\ \ref{fig_USco:completeness}).
We should mention that WFCAM was not quite in its final best 
focus mode so that those completeness limits will be
improved for the upcoming releases.
We refer the reader to \citet{lawrence06} for more details
on the observing strategy of the UKIDSS project.

%
%
\begin{table}
 \centering
  \caption{Central wavelengths, 50\% cut-on and 50\% cut-off
(in microns) for the 5 WFCAM bandpasses.
}
 \label{tab_USco:filters}
 \begin{tabular}{c c c c c c}
 \hline
   Passband:       & $Z$   & $Y$   & $J$   & $H$   & $K$  \cr
 \hline
Central wavelength & 0.88  & 1.03  & 1.25  & 1.65  & 2.20 \cr
50\% cut-on        & 0.83  & 0.97  & 1.17  & 1.49  & 2.03 \cr
50\% cut-off       & 0.925 & 1.07  & 1.33  & 1.78  & 2.37 \cr
 \hline
 \end{tabular}
\end{table}
%

%
%

All observations carried out by the UKIDSS 
project\footnote{www.ukidss.org}
are pipeline--processed at the Cambridge Astronomical Survey Unit
(CASU; Irwin et al.\ 2006, in preparation)\footnote{The CASU 
WFCAM webpage can be found at http://apm15.ast.cam.ac.uk/wfcam}.
The processed data are then archived in Edinburgh and released to the
user community through the WFCAM Science Archive
(WSA; Hambly et al.\ 2006, in preparation)\footnote{The WFCAM 
Science Archive is accessible at http://surveys.roe.ac.uk/wsa}.

We have selected all point sources (yjhkClass = $-$1 or $-$2)
in the Upper Sco association taken during the Science Verification phase. 
The Class parameter refers to a morphological classification
of a source in each passband where $-$1 is a point source,
$-$2 is borderline stellar, 0 is noise, and $+$1 is nonstellar
(more details can be found in the online glossary within the 
WFCAM Science Archive pages).
We have selected sources detected in all of $YJHK$ with no requirement
for detection in $Z$ in order to increase our sensitivity to
cooler objects. The query included the cross--correlation
with 2MASS to compute their proper motion thanks to a 5 year
baseline between the GCS and 2MASS observations in Upper Sco.
The query returned a total of 133,476 sources 
fainter than $Z$ = 11.4 and 
$J$ = 10.5 mag to avoid contamination from saturated stars.
The magnitudes are AperMag3 magnitudes and the colours originate
from the difference of those magnitudes. 
Those AperMag3 magnitudes represent the flux of a point source within
an aperture with a diameter of 1 arcsec.
The Structured Query Language query used to select the initial sample
is given in Appendix~\ref{sql}, along with an explanation.
The coverage, displayed in
Fig.\ \ref{fig_USco:coverage1}, is about 6.5 square degrees.
The resulting colour--magnitude diagram 
is shown in Fig.\ \ref{fig_USco:ZJZcmd}
and is analysed in the next section.

%
%
\begin{table}
 \centering
  \caption{Log of the observations. Central coordinates (in J2000) of each
WFCAM tile is provided along with the observing date. Each tile was observed
in all five passbands ($ZYJHK$).
}
 \label{tab_USco:log_obs}
 \begin{tabular}{c c c c c c}
 \hline
Tile  &  R.A.\  & Dec.\ & Date & \\
 \hline 
1 	& 	16 	8 	& 	--22 	40 	& 	2005-04-18 \\
2 	& 	16 	8 	& 	--23 	33 	& 	2005-04-18 \\
3 	& 	16 	9 	& 	--21 	48 	& 	2005-04-11 \\
4 	& 	16 	11 	& 	--22 	41 	& 	2005-04-11 \\
5 	& 	16 	11 	& 	--23 	33 	& 	2005-04-18 \\
6 	& 	16 	13 	& 	--21 	48 	& 	2005-04-11 \\
7 	& 	16 	15 	& 	--22 	41 	& 	2005-04-11 \\
8 	& 	16 	15 	& 	--23 	33 	& 	2005-04-18 \\
 \hline
 \end{tabular}
\end{table}

%
%
\begin{table*}
\centering
\caption[]{Summary of quality indicators for the observations, averaged over
all detector frames of a given filter. These illustrate the quality of the
science verification data, and are not typical of the UKIDSS surveys in
general \citep{dye06}. Magnitudes are on the Vega magnitude scale in the 
natural WFCAM Mauna Kea Observatories (MKO) photometric system, as defined 
in Hewett et al.~(2006) and references therein.}
\label{qc}
\begin{tabular}{c c c c c c c c c c c c c}
\hline
Filter & \multicolumn{3}{c}{Seeing / arcsec} & \multicolumn{3}{c}{Average stellar ellipticity} & \multicolumn{3}{c}{Photometric zeropoint} & \multicolumn{3}{c}{5$\sigma$ detection limit} \\
       & min & mean & max & min & mean & max & min & mean & max & min & mean & max \\
\hline
 	Z 	& 	0.80 	& 	1.02 	& 	1.41 	& 	0.036 	& 	0.111 	& 	0.239 	& 	22.68 	& 	22.72 	& 	22.76 	& 	19.91 	& 	20.31 	& 	20.79 	\\
 	Y 	& 	0.84 	& 	1.04 	& 	1.38 	& 	0.040 	& 	0.105 	& 	0.241 	& 	22.64 	& 	22.67 	& 	22.69 	& 	19.45 	& 	19.86 	& 	20.17 	\\
 	J 	& 	0.70 	& 	0.90 	& 	1.24 	& 	0.031 	& 	0.110 	& 	0.273 	& 	24.40 	& 	24.48 	& 	24.49 	& 	19.17 	& 	19.41 	& 	19.67 	\\
 	H 	& 	0.68 	& 	0.86 	& 	1.02 	& 	0.030 	& 	0.098 	& 	0.238 	& 	24.68 	& 	24.72 	& 	24.73 	& 	18.49 	& 	18.77 	& 	19.03 	\\
 	K 	& 	0.77 	& 	0.95 	& 	1.18 	& 	0.036 	& 	0.123 	& 	0.274 	& 	24.01 	& 	24.03	& 	24.04 	& 	17.74 	& 	18.02 	& 	18.18 	\\
 \hline
 \end{tabular}
\end{table*}

%
%

%
%
\section{New association members in Upper Sco}
\label{USco:cand}

This section describes the selection of cluster candidates in
Upper Sco based on various colour--magnitude diagrams drawn from
the $ZYJHK$ broadband filters and proper motion using
2MASS \citep{cutri03} as a measure of first epoch positions for
the brighter objects.

\subsection{Colour--magnitude diagrams}

\begin{figure*}
   \centering
   \includegraphics[width=0.49\linewidth]{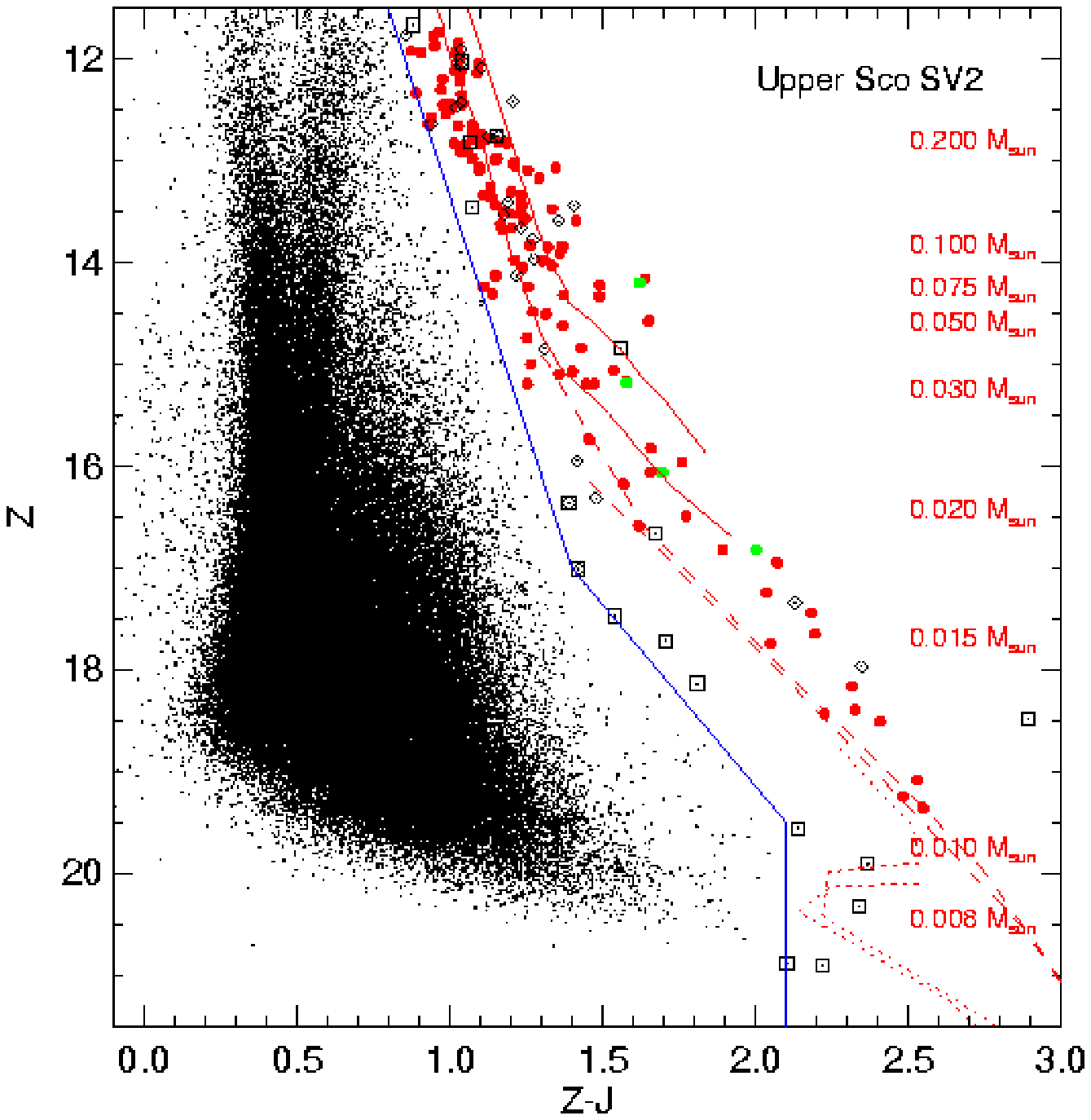}
   \includegraphics[width=0.49\linewidth]{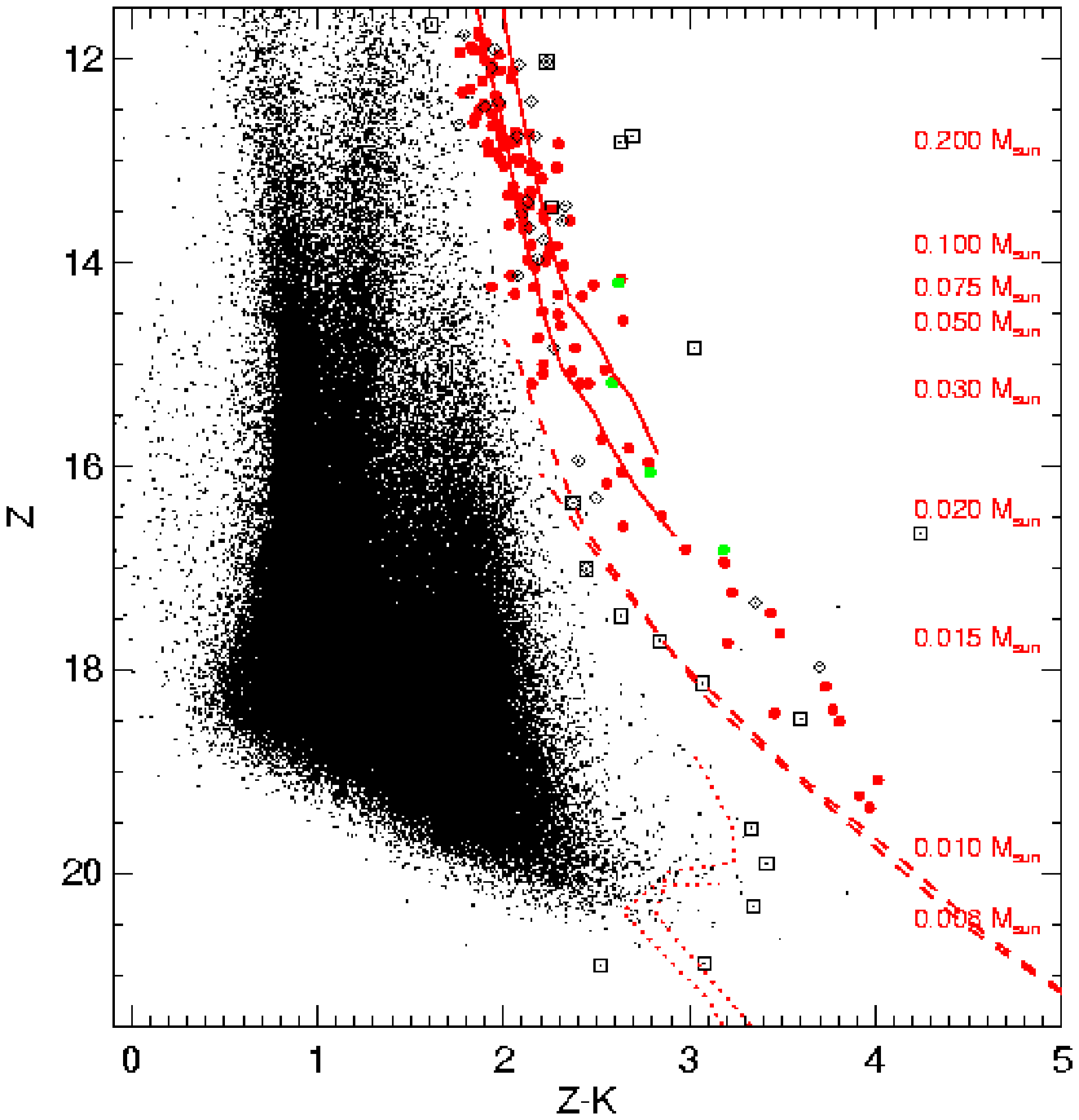}
   \includegraphics[width=0.49\linewidth]{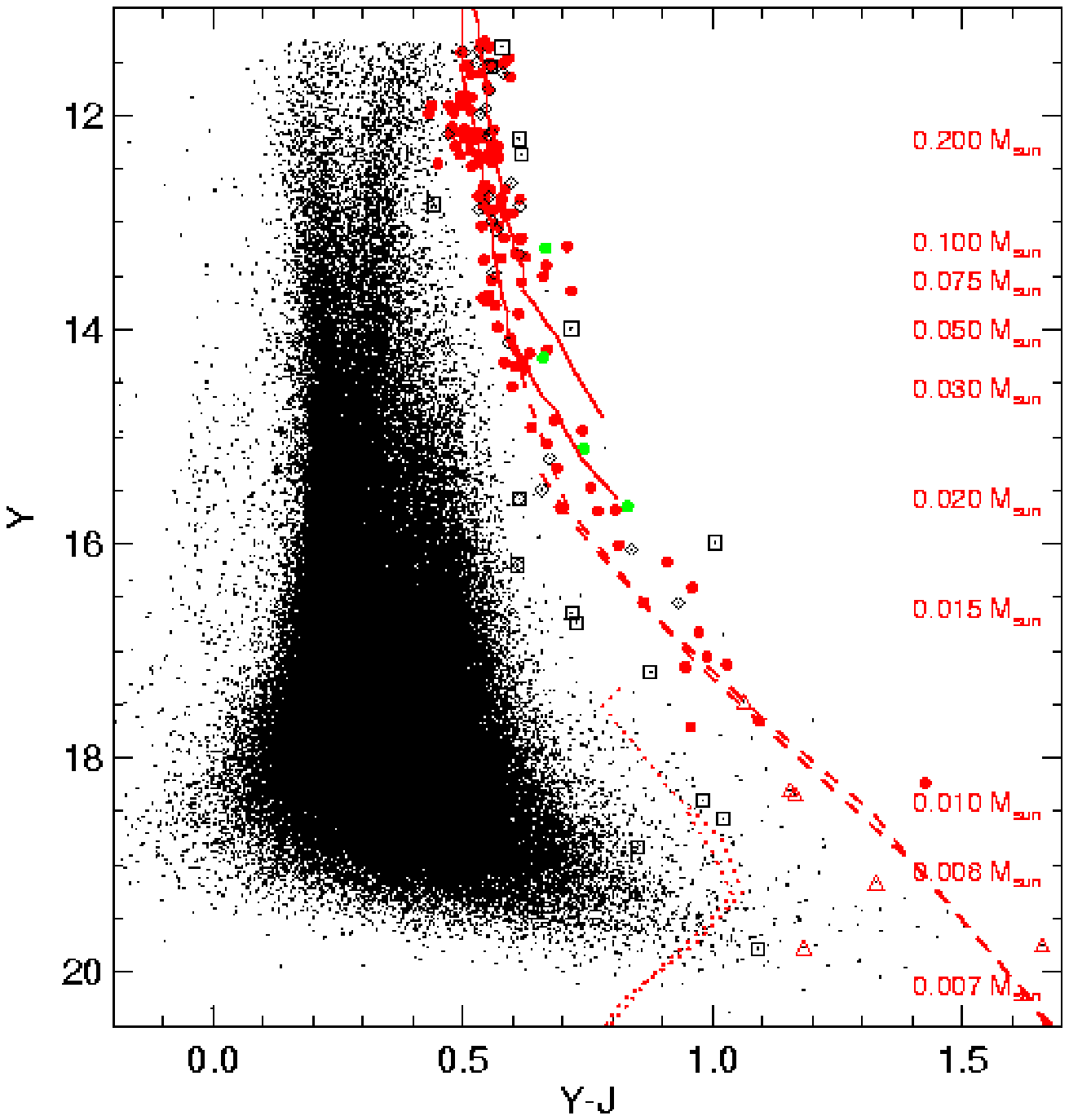}
   \includegraphics[width=0.49\linewidth]{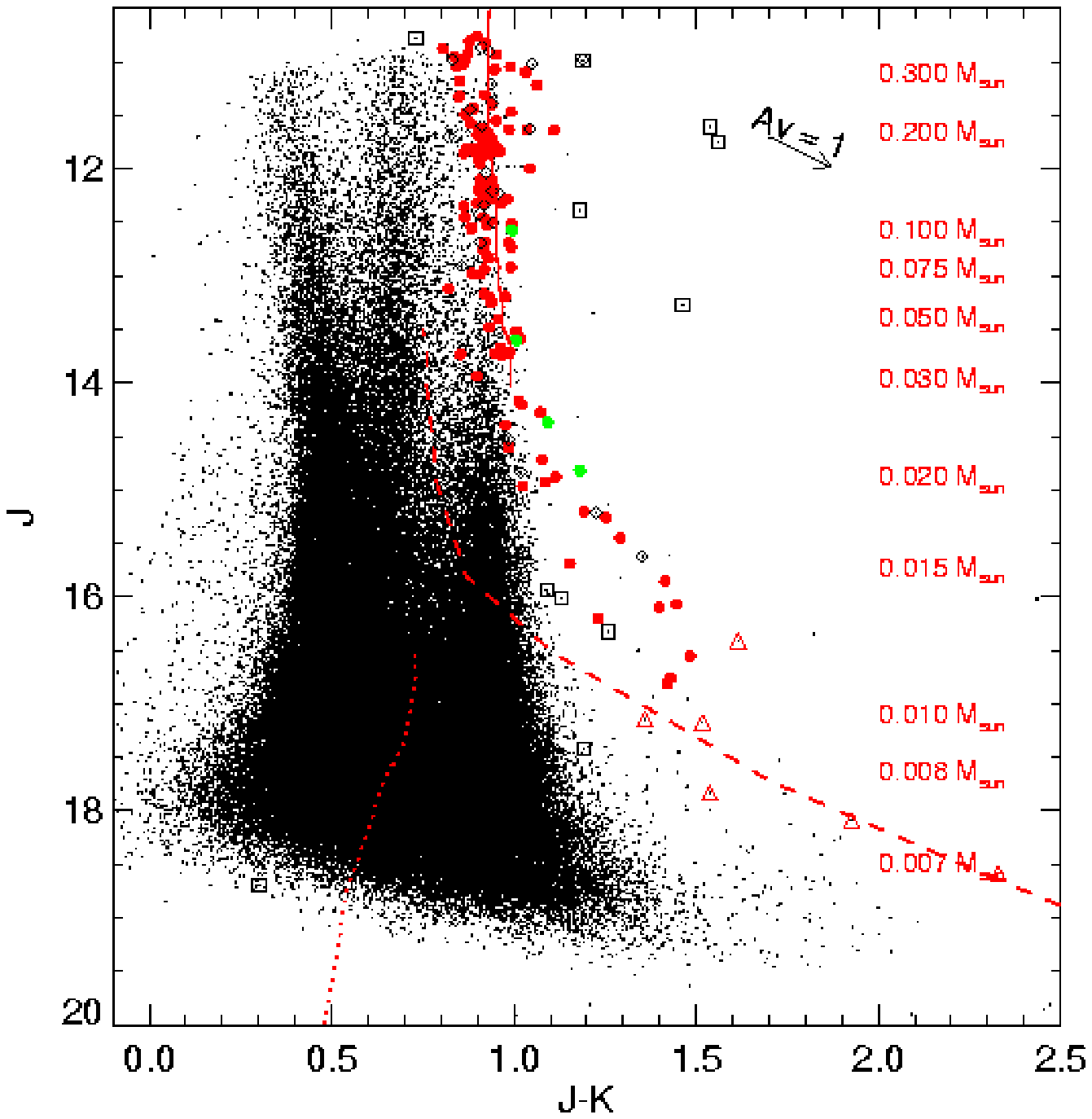}
   \caption{Colour-magnitude diagrams for 6.5
square degrees in Upper Sco from the
UKIDSS Galactic Cluster Survey.
Red circles represent the selected candidates
in Upper Sco from the GCS Science Verification\@.
Open triangles are new faint candidates with no
$Z$--band detection extracted from the ($Y-J$,$Y$)
colour--magnitude diagram.
Photometric non-members and proper motion non-members
are displayed as open squares and open diamonds, respectively.
Green circles are spectroscopic members recovered
in our study \citep{martin04,slesnick06}.
Overplotted are the 5 and 10 Myr NextGen \citep[solid line;][]{baraffe98},
DUSTY \citep[dashed line;][]{chabrier00c}, and
COND \citep[dotted line;][]{baraffe02} isochrones.
The mass scale is shown on the right hand side of the diagrams
and spans 0.3--0.01 M$_{\odot}$, according to the NextGen
and DUSTY models.
{\it{Top left:}} ($Z-J$,$Z$) diagram.
{\it{Top right:}} ($Z-K$,$Z$) diagram.
{\it{Bottom left:}} ($Y-J$,$Y$) diagram.
{\it{Bottom right:}} ($J-K$,$J$) diagram.
Note that the theoretical isochrones were specifically computed
for the WFCAM set of filters (courtesy I.\ Baraffe and F.\ Allard).
}
   \label{fig_USco:cmds}
\end{figure*}

The depth and coverage of the GCS observations provides 
$ZYJHK$ photometry of 133,476 point sources in Upper Sco 
over 6.5 square degrees with $Z$ = 11.4--21.5 mag. 
We have plotted in Fig.\ \ref{fig_USco:cmds}
four colour--magnitude diagrams to select potential cluster 
candidates and assess their membership. On the top panels
in Fig.\ \ref{fig_USco:cmds} we show the ($Z-J$,$Z$) and
($Z-K$,$Z$) diagrams whereas the ($Y-J$,$Y$) and ($J-K$,$J$) 
diagrams are displayed in the lower panels.
Overplotted are 5 and 10 Myr NextGen \citep[solid lines;][]{baraffe98},
DUSTY \citep[dashed lines;][]{chabrier00c}, and 
COND \citep[dotted lines;][]{baraffe02} isochrones shifted
at the distance of the association (d = 145 pc).

%
%
%
\begin{figure}
   \includegraphics[width=1.00\linewidth]{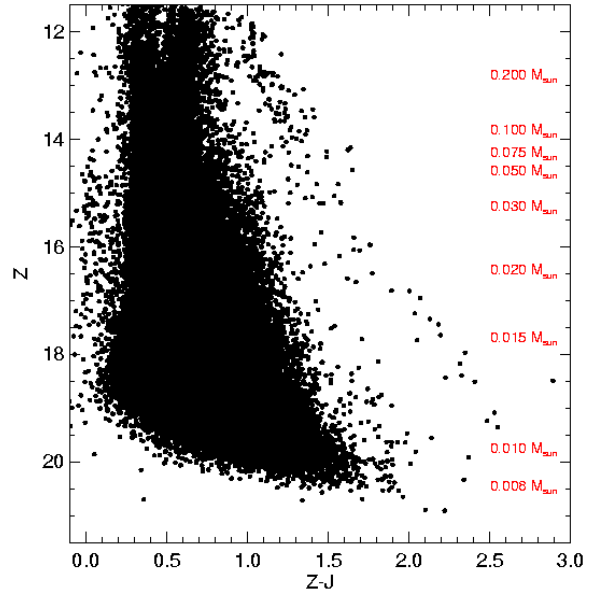}
   \caption{($Z-J$,$Z$) colour--magnitude diagram for a 6.5
square degree area covered in the Upper Sco association within the
science verification phase of the UKIDSS Galactic Cluster Survey.
The gap between the members of the association and field stars
is clearly demarcated.
The mass scale is shown on the right hand side of the diagrams
and extends below 0.01 M$_{\odot}$, according to the
DUSTY models.
}
   \label{fig_USco:ZJZcmd}
\end{figure}

%
%
\begin{figure}
   \includegraphics[width=\linewidth]{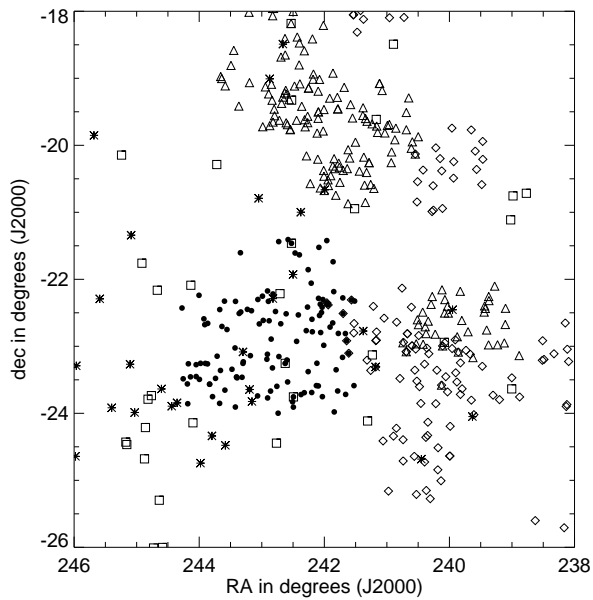}
   \caption{
            Comparison between previous and the current surveys
            conducted in the Upper Sco association. 
            New candidates from our study are shown as filled circles.
            We have included
            known members from \citet{preibisch02} (triangles),
            \citet{ardila00} (diamonds), \citet{martin04} (squares),
            and \citet{slesnick06} (star symbols).
            Four sources are common to our study and those of
            \citet{martin04} and \citet{slesnick06}.
}
   \label{fig_USco:coverage2}
\end{figure}

%
%
%

%
%
\begin{figure*}
   \centering
   \includegraphics[width=0.49\linewidth]{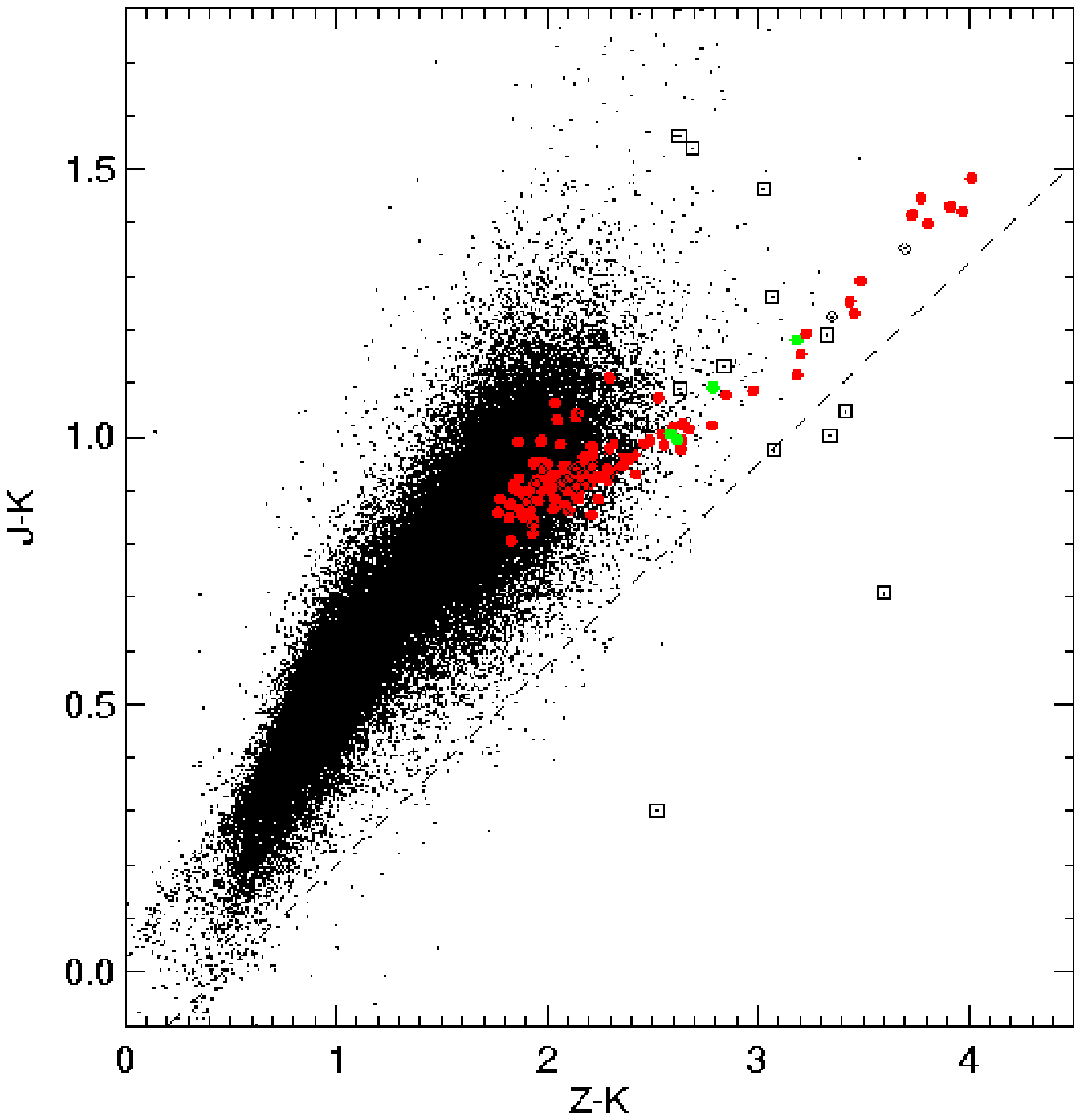}
   \includegraphics[width=0.49\linewidth]{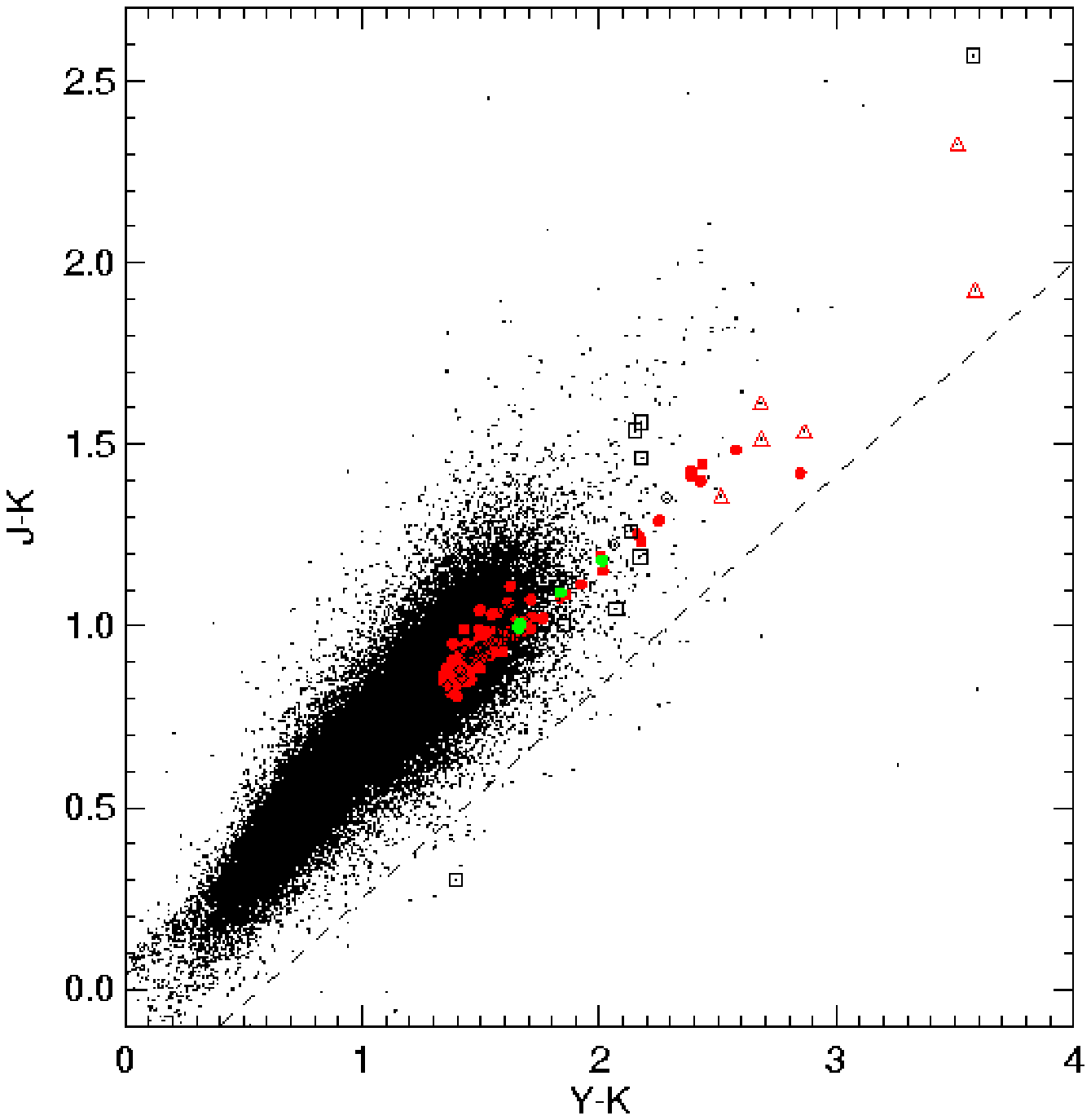}
   \caption{($Z-K$,$J-K$) and ($Y-K$,$J-K$) two--colour diagrams
for 6.5 square degrees in Upper Sco from the
UKIDSS Galactic Cluster Survey.
Red circles represent the new candidates
in Upper Sco and open triangles new faint candidates
with no $Z$--band detection.
Open squares and open diamonds are photometric and
proper motion non-members, respectively.
Green circles are spectroscopic members
\citep{martin04,slesnick06} recovered in our study.
}
   \label{fig_USco:two_col}
\end{figure*}

%
%
%
\begin{figure}
   \includegraphics[width=1.00\linewidth]{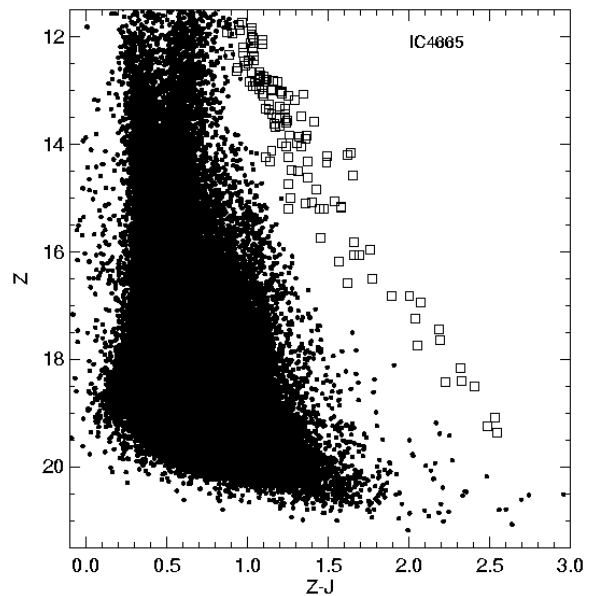}
   \caption{($Z-J$,$Z$) colour--magnitude diagram for a 2.4
square degree area covered in the IC4665 open cluster during the
science verification phase of the UKIDSS Galactic Cluster Survey.
We have superimposed the sequence of Upper Sco members (open
squares) to illustrate the low contamination of our sample.
}
   \label{fig_USco:cmdIC4665}
\end{figure}
%

%
%
\subsection{Selection of cluster candidates}
\label{USco:select}
The extraction of candidates in open clusters usually consists
in selecting sources to the right of the Zero Age Main Sequence
\citep{leggett92} or theoretical isochrones \citep{baraffe98}
shifted to the distance of the cluster.
This method has been applied at optical and near--infrared
wavelengths in numerous clusters, including the Pleiades
\citep{zapatero97a}, $\alpha$ Per \citep{barrado02a},
IC 2391 \citep{barrado01b}, and others. The contamination
by field stars along the line of sight
is estimated to about 20--40\,\% across the hydrogen burning
limit \citep{moraux01,barrado02a}.

The sequence of members in Upper Sco is well separated
from field stars in
the ($Z-J$,$Z$) colour--magnitude diagram and matches well the
theoretical isochrones over the 0.3--0.008 M$_{\odot}$ mass range
($Z$ = 11.5--20.5 mag). The masses indicated to the right hand
side of each diagram come from the NextGen models for objects
more massive than 0.05 M$_{\odot}$ and DUSTY isochrones for
lower masses, assuming an age of 5 Myr and a distance of
145 pc for the association \citep{preibisch02}.

We have selected a total of 164 sources to the right of three
segments defined by:
\begin{itemize}
\item ($Z-J$,$Z$) = (0.8,11.5) and (1.4,17.0)
\item ($Z-J$,$Z$) = (1.4,17.0) and (2.1,19.5)
\item All objects satisfying $Z-J >$2.1 and $J <$19.5 mag
\end{itemize}
We believe that our selection is conservative as
we included all cluster members and field stars lying
on the blue side of the gap between the association and field
sequences. This selection also allows for the
line--of--sight depth of
the association. Field dwarfs and background stars
contaminating our sample are excluded from the original
list after investigation of their location in other
colour--magnitude diagrams and in the proper motion vector point
diagram (see later). For example,
we found 18 sources (open squares) lying clearly either
on the blue side or the red side of the association sequence
in the other colour--magnitude diagrams displayed
in Fig.\ \ref{fig_USco:cmds} and in the ($Z-K$,$J-K$)
colour--colour diagram (Fig.\ \ref{fig_USco:two_col}).
Consequently, 146 candidates remain probable photometric
members. We further investigate their membership
in Sect.\ \ref{USco:PM} using their proper motion.
We note that our study recovered 4 spectroscopic members
(Table \ref{tab_USco:common4}; green filled circles in
Figs.\ \ref{fig_USco:cmds} \& \ref{fig_USco:two_col})
unveiled by \citet{martin04} and \citet{slesnick06}.
One of them, DENIS J160958$-$234519 was reported as
binary member of Upper Sco by \citet{bouy06b}.

We would like to emphasize that the contamination of our
sample by field dwarfs or background stars is negligible.
To corroborate this statement, we have extracted from the
WSA all point sources towards the young open cluster IC4665
(50--100 Myr; 350 pc) and found 2 contaminants after
applying the same
photometric and proper motion selection. This result represent
an upper limit on the contamination as IC4665 is at lower
galactic latitude than Upper Sco
(b $\sim$ 10$^{\circ}$ vs.\ 20$^{\circ}$), while the surveyed
area is similar.
Figure \ref{fig_USco:cmdIC4665} shows the ($Z-J$,$Z$)
colour-magnitude diagram for IC4665 with the Upper Sco
members superimposed (open squares). It is immediately
obvious from this diagram that the field contamination
must be very low.

Furthermore, we have searched for lower mass brown dwarfs
by selecting faint candidates in the ($Y-J$,$Y$)
colour-magnitude diagram with no $Z$-band detections,
yielding 15 new candidates (red open triangles in
Figs.\ \ref{fig_USco:cmds} \& \ref{fig_USco:two_col}).
The ($J-K$,$J$) diagram suggests however
that several of them are likely contaminants since
their $J-K$ colours are too blue. This leaves
6 sources worthy of spectroscopic follow--up after
retaining sources redder than $J-K$ = 1.2\@.
One object (USco J160843.4$-$224516; $J$ = 18.6; $J-K$ = 2.3)
would be the faintest and coolest brown dwarf found in Upper Sco
if confirmed as a member. Its mass and effective temperature
would be estimated as
$\sim$7 M$_{\rm Jup}$ and $\sim$1700 K by the DUSTY models
\citep{chabrier00c}. The mass estimate should be interpreted
with caution as \citet{mohanty04a} derived masses from
high-resolution spectroscopy discrepant by a factor of 2
with model predictions for the two faintest brown dwarfs
found in Upper Sco prior to our survey. Our survey
reaches three magnitudes deeper than any previous study
in the region and has uncovered a large number of new low--mass
brown dwarfs below 0.02 M$_{\odot}$.

%
%
\begin{table*}
  \caption{Near--infrared ($ZYJHK$) photometry for 129
member candidates in Upper Sco selected on the 
basis of their infrared colours. This table lists the equatorial
coordinates (at equinox J2000), the magnitudes from the GCS and their 
associated errors taken from the WSA archive, as well
as their proper motions.
Proper motions are set arbitrarily to 99.9 when unavailable (13 objects);
116 objects have proper motions consistent with cluster membership.
}
  \label{tab_USco:new_cand}
  \begin{tabular}{c c c c c c c c c}
  \hline
R.A.\ & Dec.\  &  $Z$  &  $Y$  &  $J$  &  $H$  & $K$ & $\mu_{\alpha}cos\delta$ & $\mu_{\delta}$ \\ \hline
16 06 03.75 & $-$22 19 30.0 & 18.169$\pm$0.036 & 16.825$\pm$0.014 & 15.853$\pm$0.009 & 15.096$\pm$0.009 & 14.438$\pm$0.009 &   13.7 &  $-$26.5 \\
16 06 06.29 & $-$23 35 13.3 & 18.430$\pm$0.041 & 17.150$\pm$0.018 & 16.204$\pm$0.012 & 15.540$\pm$0.012 & 14.973$\pm$0.012 &   99.9   &   99.9 \\
16 06 15.95 & $-$22 18 28.0 & 14.307$\pm$0.003 & 13.705$\pm$0.002 & 13.166$\pm$0.002 & 12.570$\pm$0.001 & 12.248$\pm$0.002 &   $-$5.0 &  $-$15.2 \\
16 06 26.37 & $-$23 06 11.4 & 14.477$\pm$0.003 & 13.770$\pm$0.002 & 13.205$\pm$0.002 & 12.616$\pm$0.001 & 12.271$\pm$0.002 &    0.7 &  $-$22.0 \\
16 06 34.61 & $-$22 55 04.4 & 13.049$\pm$0.002 & 12.366$\pm$0.001 & 11.835$\pm$0.001 & 11.243$\pm$0.001 & 10.873$\pm$0.001 &  $-$13.4 &   $-$8.8 \\
16 06 38.24 & $-$23 43 03.8 & 11.867$\pm$0.001 & 11.413$\pm$0.001 & 10.915$\pm$0.001 & 10.438$\pm$0.000 & 10.041$\pm$0.000 &  $-$13.5 &  $-$25.6 \\
16 06 39.22 & $-$22 48 34.2 & 13.520$\pm$0.002 & 12.874$\pm$0.001 & 12.322$\pm$0.001 & 11.742$\pm$0.001 & 11.422$\pm$0.001 &   $-$9.5 &  $-$19.6 \\
16 06 48.18 & $-$22 30 40.1 & 16.818$\pm$0.013 & 15.697$\pm$0.007 & 14.926$\pm$0.005 & 14.353$\pm$0.005 & 13.839$\pm$0.006 &  $-$22.7 &  $-$14.8 \\
16 06 49.10 & $-$22 16 38.4 & 15.094$\pm$0.004 & 14.339$\pm$0.003 & 13.735$\pm$0.002 & 13.199$\pm$0.002 & 12.881$\pm$0.003 &   $-$6.8 &  $-$26.5 \\
16 06 50.18 & $-$23 09 54.0 & 12.971$\pm$0.001 & 12.388$\pm$0.001 & 11.816$\pm$0.001 & 11.241$\pm$0.001 & 10.902$\pm$0.001 &  $-$22.9 &  $-$10.9 \\
16 07 06.33 & $-$22 48 28.2 & 12.644$\pm$0.001 & 12.129$\pm$0.001 & 11.567$\pm$0.001 & 10.967$\pm$0.001 & 10.686$\pm$0.001 &  $-$11.4 &  $-$21.6 \\
16 07 08.81 & $-$23 39 59.9 & 12.653$\pm$0.001 & 12.127$\pm$0.001 & 11.624$\pm$0.001 & 11.044$\pm$0.000 & 10.714$\pm$0.001 &   $-$4.7 &  $-$14.5 \\
16 07 14.79 & $-$23 21 01.2 & 19.085$\pm$0.076 & 17.649$\pm$0.029 & 16.556$\pm$0.017 & 15.829$\pm$0.017 & 15.072$\pm$0.015 &   99.9 &     99.9 \\
16 07 21.96 & $-$23 58 45.3 & 14.241$\pm$0.003 & 13.543$\pm$0.002 & 12.985$\pm$0.001 & 12.422$\pm$0.001 & 12.077$\pm$0.001 &   $-$8.3 &  $-$18.7 \\
16 07 23.82 & $-$22 11 02.0 & 17.239$\pm$0.016 & 16.015$\pm$0.009 & 15.202$\pm$0.006 & 14.564$\pm$0.004 & 14.009$\pm$0.005 &  $-$11.0 &  $-$30.7 \\
16 07 26.41 & $-$21 44 17.1 & 16.171$\pm$0.008 & 15.289$\pm$0.006 & 14.601$\pm$0.004 & 14.059$\pm$0.003 & 13.615$\pm$0.004 &  $-$13.2 &  $-$31.4 \\
16 07 27.82 & $-$22 39 04.0 & 19.358$\pm$0.100 & 18.236$\pm$0.048 & 16.810$\pm$0.022 & 16.091$\pm$0.023 & 15.388$\pm$0.023 &   99.9   &   99.9 \\
16 07 37.99 & $-$22 42 47.0 & 19.240$\pm$0.090 & 17.713$\pm$0.029 & 16.757$\pm$0.019 & 16.001$\pm$0.020 & 15.327$\pm$0.020 &   99.9   &   99.9   \\
16 07 45.21 & $-$22 22 57.6 & 13.589$\pm$0.002 & 12.935$\pm$0.001 & 12.348$\pm$0.001 & 11.811$\pm$0.001 & 11.485$\pm$0.001 &  $-$17.4 &  $-$21.1 \\
16 07 50.39 & $-$22 21 02.2 & 13.509$\pm$0.002 & 12.852$\pm$0.001 & 12.277$\pm$0.001 & 11.713$\pm$0.001 & 11.295$\pm$0.001 &  $-$13.9 &  $-$10.6 \\
16 07 50.49 & $-$21 25 20.2 & 14.029$\pm$0.003 & 13.319$\pm$0.002 & 12.693$\pm$0.001 & 12.100$\pm$0.001 & 11.707$\pm$0.001 &  $-$13.9 &  $-$16.8 \\
16 08 02.17 & $-$22 59 05.9 & 13.470$\pm$0.002 & 12.691$\pm$0.001 & 12.135$\pm$0.001 & 11.548$\pm$0.001 & 11.211$\pm$0.001 &   $-$8.8 &  $-$14.9 \\
16 08 05.54 & $-$22 18 07.1 & 11.728$\pm$0.001 & 11.306$\pm$0.001 & 10.762$\pm$0.000 & 10.244$\pm$0.000 &  9.862$\pm$0.000 &  $-$20.0 &  $-$17.3 \\
16 08 07.45 & $-$23 45 05.6 & 16.056$\pm$0.008 & 15.067$\pm$0.005 & 14.398$\pm$0.003 & 13.864$\pm$0.003 & 13.422$\pm$0.004 &   $-$7.7 &  $-$11.8 \\
16 08 08.47 & $-$22 25 00.1 & 12.572$\pm$0.001 & 12.109$\pm$0.001 & 11.631$\pm$0.001 & 10.998$\pm$0.000 & 10.726$\pm$0.001 &   $-$5.0 &  $-$22.5 \\
16 08 10.81 & $-$22 29 42.9 & 13.903$\pm$0.002 & 13.156$\pm$0.002 & 12.542$\pm$0.001 & 12.030$\pm$0.001 & 11.659$\pm$0.001 &   $-$5.9 &  $-$12.1 \\
16 08 14.00 & $-$22 47 39.4 & 12.808$\pm$0.001 & 12.173$\pm$0.001 & 11.649$\pm$0.001 & 11.029$\pm$0.001 & 10.735$\pm$0.001 &   $-$4.9 &  $-$15.0 \\
16 08 15.66 & $-$22 22 20.1 & 11.773$\pm$0.001 & 11.360$\pm$0.001 & 10.825$\pm$0.000 & 10.241$\pm$0.000 &  9.903$\pm$0.000 &   $-$8.6 &  $-$27.4 \\
16 08 18.43 & $-$22 32 25.0 & 18.507$\pm$0.042 & 17.128$\pm$0.016 & 16.099$\pm$0.010 & 15.438$\pm$0.011 & 14.700$\pm$0.010 &    99.9  &    99.9  \\
16 08 20.79 & $-$21 31 23.6 & 12.792$\pm$0.002 & 12.212$\pm$0.001 & 11.677$\pm$0.001 & 11.080$\pm$0.000 & 10.777$\pm$0.001 &   $-$7.8 &  $-$15.5 \\
16 08 22.29 & $-$22 17 03.0 & 14.312$\pm$0.003 & 13.557$\pm$0.002 & 12.939$\pm$0.001 & 12.394$\pm$0.001 & 12.020$\pm$0.001 &  $-$13.1 &  $-$11.0 \\
16 08 23.03 & $-$23 35 29.5 & 11.901$\pm$0.001 & 11.467$\pm$0.001 & 10.875$\pm$0.001 & 10.562$\pm$0.000 & 10.069$\pm$0.000 &  $-$11.4 &  $-$21.1 \\
16 08 28.47 & $-$23 15 10.4 & 17.643$\pm$0.020 & 16.408$\pm$0.010 & 15.449$\pm$0.007 & 14.776$\pm$0.006 & 14.157$\pm$0.006 &  $-$11.6 &  $-$12.6 \\
16 08 30.49 & $-$23 35 11.0 & 16.949$\pm$0.014 & 15.685$\pm$0.007 & 14.879$\pm$0.005 & 14.290$\pm$0.004 & 13.763$\pm$0.005 &   $-$4.5 &  $-$12.2 \\
16 08 43.44 & $-$22 45 16.0 & 99.999$\pm$9.999 & 19.768$\pm$0.178 & 18.585$\pm$0.097 & 17.224$\pm$0.062 & 16.257$\pm$0.051 &    99.9  &  99.9 \\
16 08 46.06 & $-$22 46 59.4 & 12.844$\pm$0.001 & 12.279$\pm$0.001 & 11.794$\pm$0.001 & 11.132$\pm$0.001 & 10.844$\pm$0.001 &  $-$12.5 &  $-$11.1 \\
16 08 47.44 & $-$22 35 47.9 & 17.738$\pm$0.022 & 16.550$\pm$0.011 & 15.688$\pm$0.007 & 15.086$\pm$0.008 & 14.534$\pm$0.009 &    0.3 &  $-$19.8 \\
16 08 48.37 & $-$23 41 21.0 & 13.833$\pm$0.002 & 13.036$\pm$0.002 & 12.463$\pm$0.001 & 11.907$\pm$0.001 & 11.545$\pm$0.001 &   $-$5.0 &  $-$20.2 \\
16 08 50.34 & $-$22 03 28.7 & 12.786$\pm$0.002 & 12.214$\pm$0.001 & 11.709$\pm$0.001 & 11.095$\pm$0.000 & 10.778$\pm$0.001 &  $-$11.1 &  $-$29.2 \\
16 09 01.98 & $-$21 51 22.7 & 15.165$\pm$0.005 & 14.221$\pm$0.003 & 13.587$\pm$0.002 & 13.035$\pm$0.001 & 12.570$\pm$0.002 &    1.0 &  $-$21.9 \\
16 09 02.01 & $-$23 22 40.3 & 12.006$\pm$0.001 & 11.530$\pm$0.001 & 10.973$\pm$0.001 & 10.475$\pm$0.000 & 10.106$\pm$0.000 &   $-$2.5 &  $-$36.1 \\
16 09 04.51 & $-$22 24 52.5 & 14.568$\pm$0.003 & 13.636$\pm$0.002 & 12.918$\pm$0.001 & 12.374$\pm$0.001 & 11.926$\pm$0.001 &  $-$11.9 &  $-$10.1 \\
16 09 09.39 & $-$22 45 59.1 & 12.441$\pm$0.001 & 11.895$\pm$0.001 & 11.398$\pm$0.001 & 10.787$\pm$0.000 & 10.457$\pm$0.001 &  $-$25.8 &  $-$12.2 \\
16 09 16.89 & $-$23 41 32.6 & 13.169$\pm$0.002 & 12.433$\pm$0.001 & 11.876$\pm$0.001 & 11.286$\pm$0.001 & 10.963$\pm$0.001 &  $-$18.9 &  $-$12.7 \\
16 09 18.69 & $-$22 29 23.7 & 99.999$\pm$9.999 & 19.744$\pm$0.143 & 18.083$\pm$0.055 & 17.061$\pm$0.047 & 16.157$\pm$0.039 &   99.9   &  99.9 \\
16 09 29.39 & $-$23 43 12.2 & 15.961$\pm$0.007 & 14.941$\pm$0.004 & 14.202$\pm$0.003 & 13.640$\pm$0.003 & 13.180$\pm$0.003 &   $-$0.8 &  $-$19.7 \\
16 09 35.75 & $-$21 38 05.8 & 11.835$\pm$0.001 & 11.360$\pm$0.001 & 10.807$\pm$0.000 & 10.337$\pm$0.000 &  9.930$\pm$0.000 &    2.4 &  $-$18.7 \\
16 09 39.68 & $-$22 31 53.9 & 12.708$\pm$0.001 & 12.163$\pm$0.001 & 11.630$\pm$0.001 & 11.001$\pm$0.000 & 10.644$\pm$0.001 &  $-$12.4 &  $-$23.6 \\
16 09 46.33 & $-$22 55 33.6 & 13.096$\pm$0.002 & 12.448$\pm$0.001 & 11.998$\pm$0.001 & 11.286$\pm$0.001 & 10.953$\pm$0.001 &  $-$18.5 &  $-$18.9 \\
16 09 52.17 & $-$21 36 27.8 & 14.154$\pm$0.003 & 13.224$\pm$0.002 & 12.515$\pm$0.001 & 11.966$\pm$0.001 & 11.521$\pm$0.001 &  $-$10.4 &  $-$14.3 \\
16 09 56.34 & $-$22 22 45.5 & 99.999$\pm$9.999 & 19.158$\pm$0.126 & 17.830$\pm$0.060 & 16.993$\pm$0.038 & 16.293$\pm$0.043 &    99.9   &  99.9  \\
16 09 58.52 & $-$23 45 18.7 & 14.195$\pm$0.003 & 13.238$\pm$0.002 & 12.574$\pm$0.001 & 12.034$\pm$0.001 & 11.578$\pm$0.001 &  $-$14.0 &  $-$21.8 \\
16 09 58.78 & $-$23 54 27.5 & 12.494$\pm$0.001 & 11.972$\pm$0.001 & 11.490$\pm$0.001 & 10.944$\pm$0.000 & 10.624$\pm$0.001 &  $-$24.6 &  $-$24.9 \\
16 10 01.84 & $-$23 49 43.4 & 12.045$\pm$0.001 & 11.504$\pm$0.001 & 10.949$\pm$0.001 & 10.498$\pm$0.000 & 10.111$\pm$0.000 &  $-$10.6 &  $-$20.7 \\
16 10 06.08 & $-$21 27 44.1 & 16.823$\pm$0.013 & 15.650$\pm$0.007 & 14.820$\pm$0.004 & 14.219$\pm$0.003 & 13.638$\pm$0.004 &   $-$0.6 &  $-$17.1 \\
16 10 19.03 & $-$21 24 25.2 & 12.784$\pm$0.002 & 12.204$\pm$0.001 & 11.665$\pm$0.001 & 11.100$\pm$0.000 & 10.735$\pm$0.001 &  $-$12.5 &  $-$18.8 \\
16 10 20.87 & $-$23 31 55.7 & 13.015$\pm$0.002 & 12.355$\pm$0.001 & 11.801$\pm$0.001 & 11.267$\pm$0.001 & 10.930$\pm$0.001 &  $-$12.8 &   $-$8.9 \\
16 10 23.44 & $-$23 12 17.7 & 12.133$\pm$0.001 & 11.636$\pm$0.001 & 11.041$\pm$0.001 & 10.678$\pm$0.000 & 10.197$\pm$0.000 &   $-$0.9 &  $-$15.5 \\
16 10 26.50 & $-$22 30 53.4 & 12.446$\pm$0.001 & 11.904$\pm$0.001 & 11.467$\pm$0.001 & 10.973$\pm$0.000 & 10.475$\pm$0.001 &    5.7 &  $-$17.7 \\
16 10 30.14 & $-$23 15 16.8 & 16.062$\pm$0.008 & 15.110$\pm$0.005 & 14.367$\pm$0.003 & 13.786$\pm$0.003 & 13.275$\pm$0.003 &   $-$1.9 &  $-$16.8 \\
16 10 47.13 & $-$22 39 49.4 & 17.442$\pm$0.020 & 16.167$\pm$0.011 & 15.258$\pm$0.007 & 14.567$\pm$0.005 & 14.006$\pm$0.006 &  $-$15.1 &  $-$23.6 \\
16 10 54.29 & $-$23 09 11.1 & 14.125$\pm$0.003 & 13.531$\pm$0.002 & 12.975$\pm$0.001 & 12.440$\pm$0.001 & 12.090$\pm$0.001 &  $-$15.1 &  $-$32.6 \\
 \hline
\end{tabular}
\end{table*}

\begin{table*}
  \begin{tabular}{c c c c c c c c c}
  \hline
R.A.\ & Dec.\  &  $Z$  &  $Y$  &  $J$  &  $H$  & $K$ & $\mu_{\alpha}cos\delta$ & $\mu_{\delta}$ \\ \hline
16 10 54.99 & $-$21 26 14.0 & 14.222$\pm$0.003 & 13.398$\pm$0.002 & 12.730$\pm$0.001 & 12.156$\pm$0.001 & 11.738$\pm$0.001 &   $-$7.8 &  $-$16.8 \\
16 10 57.28 & $-$23 59 54.1 & 14.044$\pm$0.003 & 13.346$\pm$0.002 & 12.803$\pm$0.001 & 12.212$\pm$0.001 & 11.877$\pm$0.001 &   $-$8.9 &  $-$19.8 \\
16 11 02.11 & $-$23 35 50.6 & 13.024$\pm$0.002 & 12.371$\pm$0.001 & 11.817$\pm$0.001 & 11.225$\pm$0.001 & 10.872$\pm$0.001 &  $-$27.9 &  $-$31.2 \\
16 11 07.38 & $-$22 28 50.3 & 13.585$\pm$0.002 & 12.785$\pm$0.002 & 12.171$\pm$0.001 & 11.641$\pm$0.001 & 11.226$\pm$0.001 &    0.0 &  $-$28.9 \\
16 11 17.05 & $-$22 13 08.8 & 12.826$\pm$0.002 & 12.154$\pm$0.001 & 11.639$\pm$0.001 & 10.982$\pm$0.000 & 10.528$\pm$0.001 &  $-$19.2 &  $-$26.1 \\
16 11 19.07 & $-$23 19 20.4 & 12.736$\pm$0.001 & 12.171$\pm$0.001 & 11.635$\pm$0.001 & 10.951$\pm$0.000 & 10.598$\pm$0.001 &  $-$15.0 &  $-$18.2 \\
16 11 23.99 & $-$22 53 32.6 & 12.537$\pm$0.001 & 11.981$\pm$0.001 & 11.549$\pm$0.001 & 10.937$\pm$0.000 & 10.599$\pm$0.001 &  $-$11.0 &   $-$7.2 \\
16 11 26.30 & $-$23 40 06.1 & 14.834$\pm$0.004 & 13.975$\pm$0.003 & 13.404$\pm$0.002 & 12.822$\pm$0.001 & 12.448$\pm$0.002 &   $-$1.2 &  $-$24.8 \\
16 11 31.81 & $-$22 37 08.3 & 13.984$\pm$0.003 & 13.286$\pm$0.002 & 12.679$\pm$0.001 & 12.115$\pm$0.001 & 11.758$\pm$0.001 &   $-$1.7 &  $-$16.6 \\
16 11 34.70 & $-$22 19 44.3 & 14.613$\pm$0.004 & 13.854$\pm$0.003 & 13.242$\pm$0.002 & 12.693$\pm$0.001 & 12.304$\pm$0.001 &   $-$4.3 &  $-$14.7 \\
16 11 37.61 & $-$23 46 14.8 & 13.302$\pm$0.002 & 12.643$\pm$0.001 & 12.099$\pm$0.001 & 11.478$\pm$0.001 & 11.158$\pm$0.001 &  $-$21.5 &   $-$9.6 \\
16 11 37.84 & $-$22 10 27.5 & 12.107$\pm$0.001 & 11.606$\pm$0.001 & 11.069$\pm$0.001 & 10.543$\pm$0.000 & 10.122$\pm$0.000 &   $-$4.2 &  $-$26.1 \\
16 11 38.37 & $-$23 07 07.5 & 15.190$\pm$0.004 & 14.367$\pm$0.003 & 13.743$\pm$0.002 & 13.204$\pm$0.002 & 12.777$\pm$0.002 &  $-$12.1 &  $-$31.9 \\
16 11 40.40 & $-$23 11 34.9 & 13.847$\pm$0.002 & 13.143$\pm$0.002 & 12.525$\pm$0.001 & 11.987$\pm$0.001 & 11.600$\pm$0.001 &   $-$4.7 &  $-$15.5 \\
16 11 54.39 & $-$22 36 49.3 & 15.730$\pm$0.007 & 14.912$\pm$0.005 & 14.274$\pm$0.003 & 13.628$\pm$0.002 & 13.201$\pm$0.003 &    0.4 &  $-$11.9 \\
16 11 57.37 & $-$22 15 06.8 & 15.071$\pm$0.004 & 14.287$\pm$0.003 & 13.668$\pm$0.002 & 13.095$\pm$0.001 & 12.705$\pm$0.002 &  $-$13.1 &  $-$15.9 \\
16 12 09.48 & $-$22 39 57.1 & 12.189$\pm$0.001 & 11.764$\pm$0.001 & 11.212$\pm$0.001 & 10.425$\pm$0.000 & 10.149$\pm$0.000 &   $-$7.8 &   $-$8.9 \\
16 12 14.92 & $-$22 18 04.0 & 13.500$\pm$0.002 & 12.882$\pm$0.002 & 12.320$\pm$0.001 & 11.726$\pm$0.001 & 11.398$\pm$0.001 &   $-$3.6 &  $-$16.5 \\
16 12 16.09 & $-$23 44 25.0 & 14.505$\pm$0.003 & 13.737$\pm$0.002 & 13.188$\pm$0.002 & 12.545$\pm$0.001 & 12.212$\pm$0.002 &   $-$7.3 &  $-$18.9 \\
16 12 27.64 & $-$21 56 40.8 & 99.999$\pm$9.999 & 18.296$\pm$0.050 & 17.141$\pm$0.028 & 16.371$\pm$0.020 & 15.782$\pm$0.024 &    99.9  &   99.9   \\
16 12 28.95 & $-$21 59 36.1 & 99.999$\pm$9.999 & 17.471$\pm$0.025 & 16.407$\pm$0.015 & 15.565$\pm$0.010 & 14.791$\pm$0.010 &    99.9  &   99.9  \\
16 12 39.54 & $-$22 28 08.3 & 12.348$\pm$0.001 & 11.826$\pm$0.001 & 11.308$\pm$0.001 & 10.714$\pm$0.000 & 10.389$\pm$0.001 &  $-$20.1 &  $-$16.6 \\
16 12 43.74 & $-$23 08 23.2 & 12.986$\pm$0.002 & 12.393$\pm$0.001 & 11.836$\pm$0.001 & 11.244$\pm$0.001 & 10.892$\pm$0.001 &  $-$17.5 &  $-$11.0 \\
16 12 46.92 & $-$23 38 40.9 & 15.180$\pm$0.005 & 14.261$\pm$0.003 & 13.601$\pm$0.002 & 13.022$\pm$0.002 & 12.595$\pm$0.002 &   $-$5.3 &  $-$17.1 \\
16 12 55.28 & $-$22 26 54.4 & 13.564$\pm$0.002 & 12.913$\pm$0.002 & 12.314$\pm$0.001 & 11.710$\pm$0.001 & 11.347$\pm$0.001 &   $-$6.7 &  $-$27.6 \\
16 13 02.32 & $-$21 24 28.4 & 99.999$\pm$9.999 & 18.335$\pm$0.049 & 17.169$\pm$0.023 & 16.371$\pm$0.019 & 15.652$\pm$0.022 &   99.9  &   99.9    \\
16 13 10.82 & $-$23 13 51.6 & 13.621$\pm$0.002 & 13.018$\pm$0.002 & 12.454$\pm$0.001 & 11.940$\pm$0.001 & 11.590$\pm$0.001 &   $-$7.2 &  $-$40.8 \\
16 13 12.16 & $-$23 15 16.6 & 12.782$\pm$0.001 & 12.261$\pm$0.001 & 11.702$\pm$0.001 & 11.104$\pm$0.001 & 10.791$\pm$0.001 &   $-$4.3 &  $-$25.3 \\
16 13 15.65 & $-$23 27 44.2 & 11.918$\pm$0.001 & 11.548$\pm$0.001 & 11.043$\pm$0.001 & 10.279$\pm$0.000 & 10.053$\pm$0.000 &   $-$7.4 &   $-$8.7 \\
16 13 20.53 & $-$22 29 16.0 & 11.929$\pm$0.001 & 11.531$\pm$0.001 & 11.021$\pm$0.001 & 10.578$\pm$0.000 & 10.162$\pm$0.000 &  $-$17.2 &  $-$11.1 \\
16 13 21.91 & $-$21 36 13.7 & 11.955$\pm$0.001 & 11.513$\pm$0.001 & 10.930$\pm$0.000 & 10.359$\pm$0.000 &  9.978$\pm$0.000 &   $-$6.8 &  $-$18.9 \\
16 13 26.66 & $-$22 30 35.0 & 15.053$\pm$0.004 & 14.185$\pm$0.003 & 13.515$\pm$0.002 & 12.959$\pm$0.001 & 12.509$\pm$0.002 &    5.6 &  $-$20.8 \\
16 13 34.76 & $-$23 28 15.7 & 14.738$\pm$0.004 & 14.081$\pm$0.003 & 13.485$\pm$0.002 & 12.914$\pm$0.002 & 12.553$\pm$0.002 &   $-$6.0 &  $-$16.8 \\
16 13 36.47 & $-$23 27 35.5 & 13.332$\pm$0.002 & 12.683$\pm$0.001 & 12.098$\pm$0.001 & 11.545$\pm$0.001 & 11.189$\pm$0.001 &  $-$13.5 &  $-$23.1 \\
16 13 36.88 & $-$23 27 29.9 & 12.828$\pm$0.001 & 12.327$\pm$0.001 & 11.812$\pm$0.001 & 11.191$\pm$0.001 & 10.912$\pm$0.001 &  $-$16.7 &  $-$16.7 \\
16 13 40.79 & $-$22 19 46.1 & 16.494$\pm$0.010 & 15.477$\pm$0.007 & 14.721$\pm$0.004 & 14.153$\pm$0.003 & 13.643$\pm$0.004 &  $-$14.4 &  $-$18.1 \\
16 13 41.30 & $-$23 54 22.0 & 13.048$\pm$0.002 & 12.470$\pm$0.001 & 11.951$\pm$0.001 & 11.377$\pm$0.001 & 11.043$\pm$0.001 &  $-$26.4 &  $-$24.2 \\
16 13 42.64 & $-$23 01 28.0 & 15.189$\pm$0.005 & 14.337$\pm$0.003 & 13.718$\pm$0.002 & 13.149$\pm$0.002 & 12.730$\pm$0.002 &  $-$13.1 &  $-$13.8 \\
16 13 54.34 & $-$23 20 34.4 & 12.114$\pm$0.001 & 11.611$\pm$0.001 & 11.095$\pm$0.001 & 10.493$\pm$0.000 & 10.063$\pm$0.000 &   $-$8.0 &  $-$24.7 \\
16 14 13.52 & $-$22 44 58.0 & 13.434$\pm$0.002 & 12.825$\pm$0.002 & 12.283$\pm$0.001 & 11.680$\pm$0.001 & 11.344$\pm$0.001 &   $-$7.0 &  $-$20.7 \\
16 14 21.44 & $-$23 39 14.8 & 16.587$\pm$0.010 & 15.663$\pm$0.006 & 14.968$\pm$0.005 & 14.445$\pm$0.005 & 13.944$\pm$0.006 &    0.2 &  $-$18.0 \\
16 14 23.12 & $-$22 19 33.9 & 13.064$\pm$0.002 & 12.288$\pm$0.001 & 11.717$\pm$0.001 & 11.186$\pm$0.001 & 10.778$\pm$0.001 &  $-$10.1 &  $-$11.2 \\
16 14 32.87 & $-$22 42 13.5 & 15.823$\pm$0.007 & 14.846$\pm$0.005 & 14.163$\pm$0.003 & 13.633$\pm$0.002 & 13.149$\pm$0.003 &  $-$10.6 &  $-$18.9 \\
16 14 38.46 & $-$23 21 37.3 & 12.327$\pm$0.001 & 11.909$\pm$0.001 & 11.436$\pm$0.001 & 10.860$\pm$0.000 & 10.551$\pm$0.001 &  $-$16.0 &  $-$16.0 \\
16 14 41.19 & $-$22 27 05.4 & 12.745$\pm$0.001 & 12.149$\pm$0.001 & 11.664$\pm$0.001 & 11.021$\pm$0.000 & 10.757$\pm$0.001 &  $-$13.5 &  $-$27.5 \\
16 14 41.68 & $-$23 51 05.9 & 18.395$\pm$0.040 & 17.059$\pm$0.017 & 16.069$\pm$0.010 & 15.337$\pm$0.010 & 14.623$\pm$0.010 &     99.9  &   99.9  \\
16 14 51.31 & $-$23 08 51.7 & 13.670$\pm$0.002 & 13.036$\pm$0.002 & 12.497$\pm$0.001 & 11.895$\pm$0.001 & 11.565$\pm$0.001 &   $-$8.5 &  $-$22.2 \\
16 14 57.50 & $-$23 28 42.8 & 12.626$\pm$0.001 & 12.173$\pm$0.001 & 11.695$\pm$0.001 & 11.073$\pm$0.001 & 10.787$\pm$0.001 &  $-$18.1 &  $-$12.6 \\
16 15 08.92 & $-$23 45 04.9 & 13.360$\pm$0.002 & 12.755$\pm$0.001 & 12.224$\pm$0.001 & 11.620$\pm$0.001 & 11.280$\pm$0.001 &   $-$9.9 &  $-$16.4 \\
16 15 20.10 & $-$23 33 54.7 & 14.329$\pm$0.003 & 13.500$\pm$0.002 & 12.838$\pm$0.001 & 12.344$\pm$0.001 & 11.907$\pm$0.001 &    7.6 &  $-$29.7 \\
16 15 20.24 & $-$23 33 59.0 & 12.910$\pm$0.002 & 12.367$\pm$0.001 & 11.873$\pm$0.001 & 11.266$\pm$0.001 & 10.955$\pm$0.001 &  $-$25.2 &  $-$16.9 \\
16 15 27.43 & $-$22 39 27.7 & 12.294$\pm$0.001 & 11.818$\pm$0.001 & 11.321$\pm$0.001 & 10.764$\pm$0.000 & 10.472$\pm$0.001 &  $-$14.1 &  $-$28.2 \\
16 15 28.19 & $-$23 15 44.1 & 14.236$\pm$0.003 & 13.675$\pm$0.002 & 13.123$\pm$0.001 & 12.634$\pm$0.001 & 12.302$\pm$0.002 &  $-$12.8 &  $-$23.8 \\
16 15 36.48 & $-$23 15 17.6 & 15.190$\pm$0.005 & 14.533$\pm$0.003 & 13.935$\pm$0.002 & 13.396$\pm$0.002 & 13.037$\pm$0.003 &   $-$1.6 &  $-$15.7 \\
16 15 38.66 & $-$22 40 37.3 & 13.826$\pm$0.002 & 13.144$\pm$0.002 & 12.562$\pm$0.001 & 12.022$\pm$0.001 & 11.677$\pm$0.001 &  $-$12.2 &  $-$31.8 \\
16 15 42.06 & $-$22 35 24.0 & 12.440$\pm$0.001 & 11.952$\pm$0.001 & 11.436$\pm$0.001 & 10.842$\pm$0.000 & 10.546$\pm$0.001 &    5.5 &  $-$20.8 \\
16 15 54.87 & $-$23 15 14.8 & 12.214$\pm$0.001 & 11.728$\pm$0.001 & 11.180$\pm$0.001 & 10.645$\pm$0.000 & 10.328$\pm$0.001 &  $-$21.0 &  $-$27.0 \\
16 15 59.26 & $-$23 29 36.5 & 13.333$\pm$0.002 & 12.765$\pm$0.001 & 12.221$\pm$0.001 & 11.631$\pm$0.001 & 11.309$\pm$0.001 &  $-$12.8 &  $-$17.8 \\
16 16 00.81 & $-$22 14 19.4 & 13.089$\pm$0.002 & 12.399$\pm$0.001 & 11.834$\pm$0.001 & 11.289$\pm$0.001 & 10.943$\pm$0.001 &    1.2 &  $-$12.8 \\
16 16 11.72 & $-$23 27 05.2 & 12.973$\pm$0.002 & 12.430$\pm$0.001 & 11.898$\pm$0.001 & 11.294$\pm$0.001 & 10.996$\pm$0.001 &  $-$15.8 &  $-$14.6 \\
16 16 11.83 & $-$23 16 26.9 & 13.971$\pm$0.002 & 13.335$\pm$0.002 & 12.758$\pm$0.001 & 12.180$\pm$0.001 & 11.843$\pm$0.001 &  $-$15.5 &  $-$11.0 \\
16 16 33.43 & $-$23 27 21.2 & 13.431$\pm$0.002 & 12.774$\pm$0.001 & 12.195$\pm$0.001 & 11.669$\pm$0.001 & 11.293$\pm$0.001 &  $-$13.1 &  $-$24.9 \\
16 16 41.63 & $-$23 15 39.1 & 12.911$\pm$0.001 & 12.395$\pm$0.001 & 11.854$\pm$0.001 & 11.287$\pm$0.001 & 10.992$\pm$0.001 &  $-$19.6 &  $-$15.6 \\
16 16 43.96 & $-$23 51 25.9 & 12.830$\pm$0.001 & 12.243$\pm$0.001 & 11.728$\pm$0.001 & 11.109$\pm$0.001 & 10.800$\pm$0.001 &   $-$3.6 &  $-$25.7 \\
16 16 45.39 & $-$23 33 41.6 & 14.992$\pm$0.004 & 14.310$\pm$0.003 & 13.726$\pm$0.002 & 13.180$\pm$0.002 & 12.779$\pm$0.002 &  $-$17.2 &  $-$30.3 \\
16 17 01.47 & $-$23 29 06.0 & 13.656$\pm$0.002 & 13.025$\pm$0.002 & 12.453$\pm$0.001 & 11.887$\pm$0.001 & 11.537$\pm$0.001 &  $-$17.9 &  $-$19.6 \\
16 17 06.06 & $-$22 25 41.6 & 13.254$\pm$0.002 & 12.663$\pm$0.002 & 12.120$\pm$0.001 & 11.516$\pm$0.001 & 11.202$\pm$0.001 &  $-$15.0 &  $-$26.4 \\
 \hline
\end{tabular}
\end{table*}

%
%
\begin{table*}
 \centering
 \caption{The four candidates with $I$ magnitudes
and spectral types already published recovered during
our work. We list their coordinates (J2000), $I$ magnitudes
from \citet{martin04} and \citet{slesnick06}, $ZYJHK$
photometry from the GCS, proper motion, as well as
H$\alpha$ equivalent widths (\AA{}) and spectral types
taken from the literature.
Note that DENIS-P J161030$-$231517 was rejected as a member
by \citet{martin04} although we find its photometry and proper
motion consistent with membership.
}
 \label{tab_USco:common4}
 \begin{tabular}{c c c c c c c c c c c c}
 \hline
RA          &  dec         &  $I$  &  $Z$   &  $Y$   &  $J$   &  $H$   &   $K$ &  $\mu_{\alpha}cos{\delta}$  &  $\mu_{\delta}$ &  EW(H$\alpha$) &  SpT \\
 \hline
16 12 46.92 & $-$23 38 40.9 & 15.90 & 15.179 & 14.261 & 13.600 & 13.022 &  12.594 &  $-$5.3 &  $-$17.1 &  $-$14.7 & M6.0 \\
16 09 58.50 & $-$23 45 18.6 & 15.04 & 14.195 & 13.238 & 12.573 & 12.034 &  11.578 & $-$14.0 &  $-$21.8 &  $-$20.0 & M6.5 \\
16 10 30.10 & $-$23 15 16.7 & 17.04 & 16.062 & 15.109 & 14.367 & 13.786 &  13.275 &  $-$2.0 &  $-$16.8 &  $-$12.0 & M7.5 \\
16 10 06.00 & $-$21 27 44.6 & 18.11 & 16.822 & 15.649 & 14.819 & 14.219 &  13.638 &  $-$0.6 &  $-$17.1 &  $-$17.0 & M8.5 \\
 \hline
 \end{tabular}
\end{table*}

%
%
\subsection{Proper motions}
\label{USco:PM}

To further assess the membership of the brightest ($J \leq$15.8)
candidates, we have cross--correlated the 2MASS database with the
GCS observations to extract proper motions. The standard pipeline
astrometric reductions of WFCAM data (Irwin et al.~2006, in
preparation) use the 2MASS point source catalogue to provide
secondary astrometric standards. Hence, on scale lengths of the order
of the WFCAM detector size ($\sim10$~arcmin) all positional systematic
errors are mapped out with respect to 2MASS and a straight--forward
difference between WFCAM and 2MASS epoch positions, divided by the
epoch difference, yields {\em relative} proper motions.
Proper motions derived using such a method are relative in the sense
that the majority of stars in each field are moving slowly enough
to be assumed to be stationary, and therefore define a zeropoint
reference frame with respect to which peculiar cluster motions can
be measured. Note, however, that such proper motions are in no sense
absolute and are not on the inertial system defined by
{\em Hipparcos}, regardless of the absolute positions being ultimately
tied to the International Coordinate Reference System via the Tycho--2
catalogue reductions applied to 2MASS. Nonetheless, within the errors
to which the proper motions are determined, they are comparable to
absolute proper motions of high mass cluster members determined by
{\em Hipparcos} since the underlying assumption of approximately
zero net background stellar proper motions in these fields is valid.

\begin{figure}
   \centering
   \includegraphics[width=\linewidth]{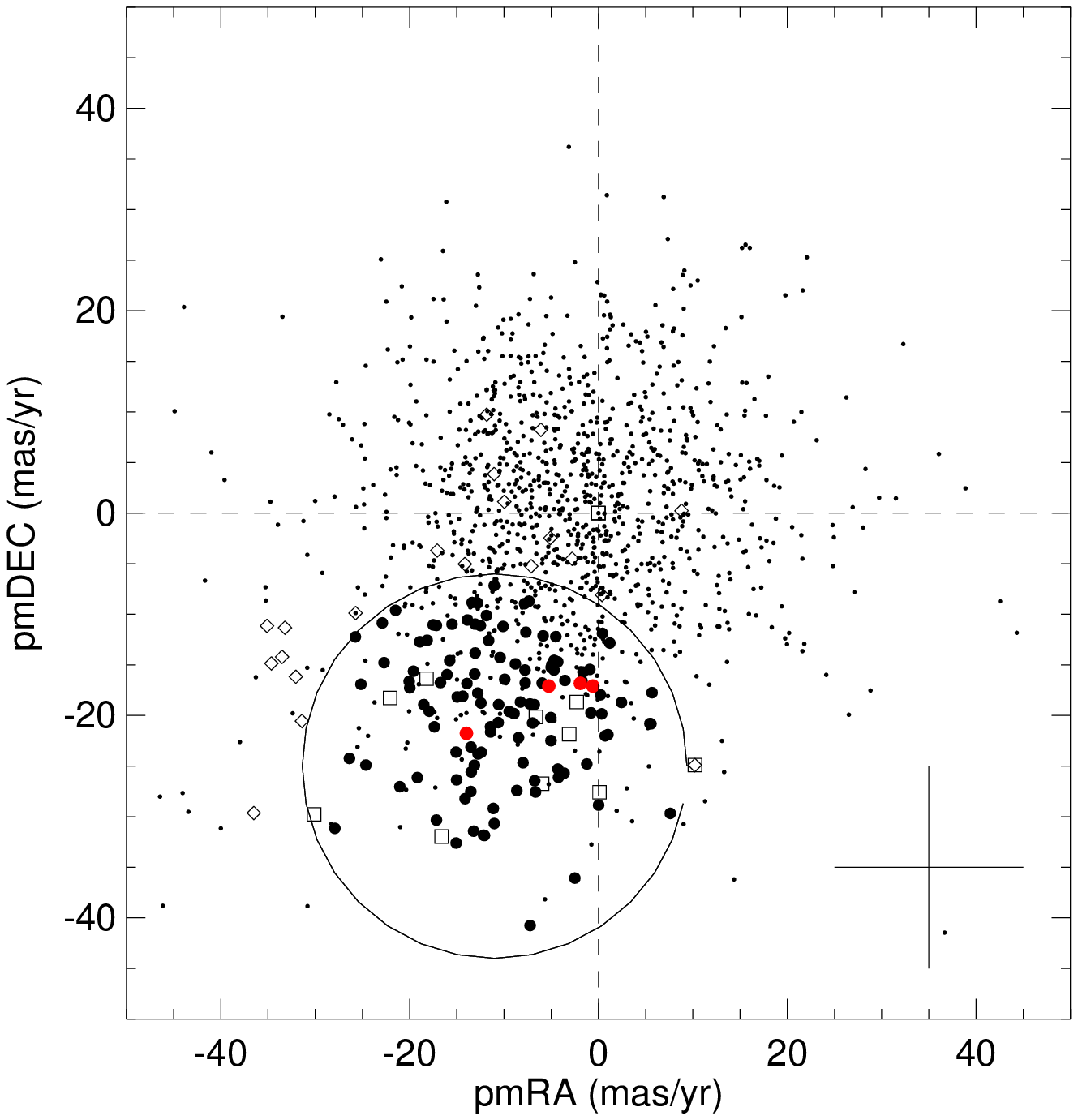}
   \caption{Vector point diagram (proper motion in right ascension
versus proper motion in declination) for photometric
candidate members in Upper Sco brighter than $J$ = 15.8 (filled circles).
We have classified as non-members candidates lying outside a
2$\sigma$ circle centred on ($-$11,$-$25) mas/yr, the mean
proper motion of Upper Sco \citep{deBruijne97,preibisch98}.
Photometric non-members and proper motion non-members
are displayed as open squares and open diamonds, respectively.
Small dots are field stars located in a 0.25 square region
(ra=242.0--242.5$^{\circ}$ and dec between $-$23.0 and $-$22.5$^{\circ}$)
to illustrate the distribution of sources in this diagram
when no photometric selection is applied.
The four red filled circles represent spectroscopic members from
\citet{martin04} and \citet{slesnick06}.
A typical proper motion error bar of 10 mas/yr is shown
in the lower right hand side.
}
   \label{fig_USco:VPM}
\end{figure}

The resulting vector point diagram (proper motion in right ascension
versus proper motion in declination) for candidates brighter
than $J$ = 15.8 mag (corresponding to 15 M$_{\rm Jup}$ according
to the DUSTY models) is shown in Fig.\ \ref{fig_USco:VPM}.
In order to assess the likely errors in the proper motions so derived,
we examined histograms of the distribution of proper motions for all
stars matched between the GCS Upper Sco observations and 2MASS. Under
the assumption that these distributions are dominated by centroiding
errors and not real stellar motions, we estimate that the proper motion
errors are $\sim10$ mas~yr$^{-1}$ in either coordinate, and we plot
a typical error bar in Fig.~\ref{fig_USco:VPM}, in addition to a
$2\sigma$ error selection circle about the
mean association proper motion of ($-$11,$-$25) mas/yr
\citep{deBruijne97,preibisch98}.
Most photometric candidates lie within the selection region
and those candidates have thus high probability of being members
of Upper Sco. We have classified 23 photometric candidates
as proper motion
non-members (open diamonds in Fig.\ \ref{fig_USco:VPM})
because they lie outside the $2\sigma$ selection
circle.
For comparison, we have also overplotted a sample
of field stars brighter than $J$ = 15.8 mag for which
proper motion is available from the GCS/2MASS cross-correlation.
We have selected them in the region delineated by
ra=242.0--242.5$^{\circ}$ and declination between
$-$23.0 and $-$22.5$^{\circ}$ (small dots in Fig.\ \ref{fig_USco:VPM}).
Clearly, candidates can not be selected on their proper
motion alone. However, proper motion does act as a useful
extra criterion to the photometry.

To summarise, we have extracted a total of 129
member candidates in Upper Sco (red symbols in
Fig.~\ref{fig_USco:VPM}) from the GCS Science Verification data
as follows. An initial sample of 164 candidates was selected
from the Z,~Z--J colour--magnitude diagram; 18 of those were
rejected on tha basis of their position in other CMDs, leaving
146 sources. A further 23 sources were rejected as having proper
motions inconsistent with cluster membership, leaving 123
candidates. Finally, we added on 6 $Z$ `drop outs' from
Y,~Y--J CMD selection, yielding a total of 129 candidates
(but note that the faintest 13 have no proper motion measurement
as they are beyond the sensitivity limit of 2MASS).

%
%

%
%
\section{Discussion}
\label{USco:discuss}
\subsection{New low--mass brown dwarfs}
\label{USco_newBDs}

The lowest mass brown dwarfs discovered in Upper Sco were reported
by \citet{martin04} with spectral types of M9\@. The faintest
of them, DENIS-P J161452$-$201713 has $J$ = 15.33 mag. Our survey
is 100\% complete, within the area surveyed,
down to $J$ = 18.7 mag i.e.\ over 3 magnitudes deeper 
than the faintest objects currently known in the association.
We detect a dozen new brown dwarf candidates
with masses below 0.02 M$_{\odot}$ (filled circles and open
triangles in Fig.\ \ref{fig_USco:cmds}), according to the DUSTY
models \citep{chabrier00c}. These new brown dwarfs are the lowest
mass substellar objects ever found in Upper Sco.
However, \citet{mohanty04b} derived
lower masses than the model predictions for two M7--M8 dwarfs
reported by \citet{ardila00} from higher resolution optical
spectra, suggesting that we may be probing lower masses
than these theoretical predictions indicate. Effective temperatures
predicted by atmospheric models suggest that the lowest mass
objects will be young L dwarfs. We plan to obtain
optical and near--infrared spectroscopy to investigate the 
effect of gravity on the absorption bands and atomic features
present in L dwarfs \citep{mcgovern04}.

We possibly detect the M7/M8 gap around 0.03 M$_{\odot}$
due to the formation of large dust grains at low temperature
\citep{dobbie02b}. This gap is consistent with the spectral types
derived for the four members recovered by our study
(green circles in Fig.\ \ref{fig_USco:cmds}; 
see also Sect.\ \ref{USco:binary}).
Finally, there is a hint of a second gap below 0.015 M$_{\odot}$
which might represent the M/L transition. Statistics on those
gaps will improve after full coverage of the association and we 
plan to investigate those effects in later papers exploiting
the UKIDSS Galactic Clusters Survey.

\subsection{Binaries}
\label{USco:binary}

Our sample of candidates in Upper Sco includes four objects
previously reported in the literature (green filled circles in
Fig.\ \ref{fig_USco:cmds}). Three of them,
DENIS-P J160958$-$234518, DENIS-P J161030$-$231516, 
and DENIS-P J161006$-$212744 were classified as M6.5, M7.5, 
and M7.5, respectively \citep{martin04}. The latter, however, was
rejected as a member on the basis of the equivalent width of the
Na{\small{I}} doublet (7.6\AA{}). This object could be a field dwarf
contaminant although it fits well the sequence of members 
in all colour--magnitude diagrams as well as in the
proper motion vector point diagram.
Radial velocity is required to further clarify the status of this object. 
The fourth object recovered, SCH161247$-$233841, was classified 
as a M6 member by \citet{slesnick06} who measured an
H$\alpha$ equivalent width of $-$14.7\AA{}.

We note that DENIS J160958$-$234518
lies clearly above the cluster sequence in the ($Z-J$,$Z$)
colour--magnitude diagram and was announced as a close
binary system at a separation of 0.08 arcsec by \citet{bouy06b}.
Other red objects lying above the sequence in the ($Z-J$,$Z$)
colour--magnitude diagram and redder in the ($J-K$,$J$)
diagram are good multiple system candidates.
The study of the binary frequency across the magnitude
range probed by our survey would shed light on the
the possible population of wide low--mass
binaries in Upper Sco recently claimed by \citet{bouy06b}.

Visual inspection of the images revealed two close systems,
USco J161520.1$-$233355 and USco J161336.9$-$232730
separated by 5.08 and 7.92 arcsec, respectively. 
Unfortunately our proper motion accuracy is not sufficient to make any
firm statement as to the physical association of either of these
potential wide binaries, but if confirmed as physical binary members 
of Upper Sco, either system would be a wide 
low--mass ($\sim$0.2 M$_{\odot}$) binary with a separation
of $\sim$1000~AU\@. Other wide low--mass binaries have recently been
reported by \citet{bouy06b} and \cite{luhman05}.
This type of binary is important to studies investigating
the formation of low--mass stars and brown dwarfs in open clusters
as well as their evolution with time.

\subsection{The IMF in Upper Sco}
\label{USco:IMF}

\begin{figure}
   \includegraphics[width=\linewidth]{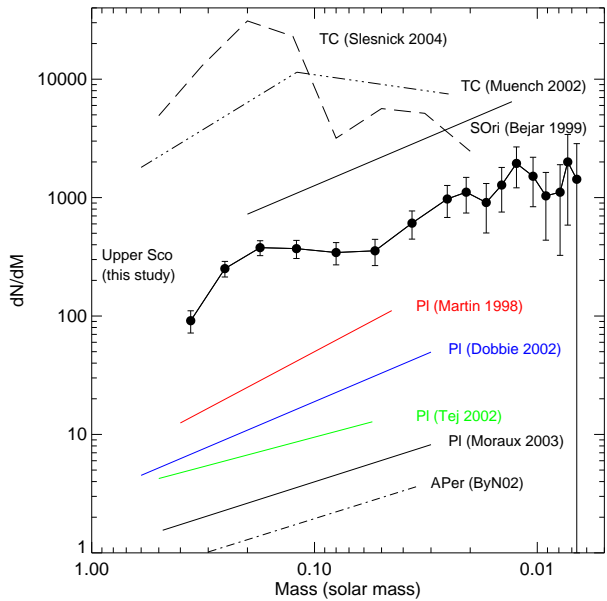}
   \caption{Cluster Initial Mass Functions for the Upper Sco
association using the members selected from their photometry
We have employed three methods to derive the Upper Sco
mass function: open triangles
represent a two-fold increase in mass, starting with
the 5--10 Jupiter masses bin. Open squares symbolise the
mass function drawn from the number of sources in 1.5
magnitude bins. Finally, filled circles correspond to
a ``smoothed'' mass function by counting the number of
objects per magnitude bin by steps of 0.5 magnitudes.
Overplotted are mass functions from the literature for
the Pleiades \citep{martin98a,tej02,dobbie02a,moraux03},
$\alpha$ Per \citep{barrado02a}, $\sigma$ Orionis \citep{bejar01},
and the Trapezium Cluster \citep{muench02,slesnick04}.
}
   \label{fig_USco:MF}
\end{figure}

We considered the 129 candidates selected in the 6.5 square degree area
surveyed in Upper Sco to derive the mass function. We have transformed
the $J$ magnitudes into masses using the NextGen models for masses above 
0.05 M$_{\odot}$ and DUSTY isochrones below \citep{baraffe98,chabrier00c}.
The number of candidates per unit of mass (dN/dM) was obtained by dividing
the number of objects in each magnitude bin (dN) by the difference of the upper
and lower limits of the bin in mass (M$_{2}$--M$_{1}$). We have derived 
the IMF, expressed as the mass spectrum (dN/dM$\propto$M$^{\alpha}$), 
using 3 approaches (we did not correct for binaries):
\begin{enumerate}
\item We counted the number of objects per mass bin starting 
at 5--10 Jupiter masses and increased it two--fold every time 
(open triangles in Fig.\ \ref{fig_USco:MF}).
\item We summed the number of candidates per 1.5 magnitude bin from
 $J$ = 11.5 mag (open squares in Fig.\ \ref{fig_USco:MF}).
\item We made a ``smoothed'' mass function by counting the number of 
candidates per magnitude bin by steps of 0.5 magnitudes, starting 
at $J$ = 10.5--11.5 mag (filled circles in Fig.\ \ref{fig_USco:MF}).
\end{enumerate}
%

%
%

The higher mass bin is likely incomplete due to our saturation 
limit occuring at $J$ = 10.5 mag, corresponding to 
M = 0.48 M$_{\odot}$. Similarly, the lower mass
bins below 0.01 M$_{\odot}$ represent lower limits because 
our survey is limited
by the $Y$ completeness ($Y$ = 19.8 mag or M $\sim$ 0.01 M$_{\odot}$). 
The best fit of the mass spectrum in Upper Sco is obtained for 
a single segment of power law index $\alpha$ = 0.6$\pm$0.1
over the 0.3--0.01 M$_{\odot}$ mass range.
The mass function is approximately linear across the entire 
mass spectrum probed by our study (Fig.\ \ref{fig_USco:MF}) 
and in agreement with estimated IMF for the Pleiades 
\citep[$\alpha$ = 0.6--1.0;][]{martin98a,tej02,dobbie02a,moraux03},
$\alpha$ Per \citep[0.59$\pm$0.05;][]{barrado02a}, 
and $\sigma$ Orionis \citep[0.8$\pm$0.4;][]{bejar01}.
Optical surveys in open clusters derived contamination 
level on the order of 30--45\% \citep{barrado02a}, even 
in proper motion samples \citep{moraux01}. We expect our
sample of candidates to be less affected by contaminants as
very few sources were detected to the right of our selection 
criteria in a control field towards the open cluster IC4665\@.
Indeed, the gap between the association sequence and field 
stars is large, particularly at fainter magnitudes.
Spectroscopic follow--up will nevertheless be required to 
verify the low level of contaminantion in order
to check our mass function in the low--mass and
substellar regimes.

Finally, we compare the mass functions of Upper Sco and
the Trapezium Cluster in Fig.~\ref{fig_USco:MF}. 
\citet{muench02} modelled the IMF from a
near--infrared survey of the Orion region. More recently,
\citet{slesnick04} obtained optical and infrared spectra of
candidates in the inner region of the Trapezium Cluster, 
yielding mass and extinction for each individual source. 
The Trapezium Cluster appears to have a break at about
0.1 M$_{\odot}$, suggesting a lower number of brown dwarfs
compared to low--mass stars in contrast to our survey in
Upper Sco. If real, this effect could be explained by the
continuous formation of brown dwarfs with time up to
5 Myr and perhaps beyond. There is no evidence for this 
hypothesis in theoretical simulations \citep{bate05,jappsen05}.
Photo--evaporation of low--mass stars could be playing some 
role in the Trapezium Cluster although very unlikely
\citep{kroupa03a}. The inner region of the Trapezium Cluster 
might also have suffered from dynamical ejection of the
lowest mass members due to the O stars.
The reason for the discrepancy in the mass functions remains 
unclear but it is too early to claim any difference due to the
uncertainties in the pre--main sequence tracks at such early
ages \citep{baraffe02}.

%
%
\section{Conclusions}
\label{USco:concl}

We have presented a deep near--infrared survey in the Upper Sco
association conducted during the Science Verification phase
of the UKIDSS Galactic Cluster Survey. The survey is complete
down to $Z$ = 20.1 mag, corresponding to 0.01 M$_{\odot}$,
and is 3 magnitudes deeper than any previous study in the
region. We have confirmed the photometric membership of
129 candidates in the 0.3--0.007 M$_{\odot}$ mass range.
Among them, 116 have proper motion consistent with
the mean motion of the association. We have discovered
a dozen new brown dwarf candidates below 15 Jupiter masses, 
according to theoretical isochrones at 5 Myr.
Finally, we have derived the mass spectrum across 
the stellar/substellar regime, yielding a power law 
with a slope $\alpha$ of 0.6$\pm$0.1, consistent with
previous studies in open clusters.

We have demonstrated the power of the UKIDSS Galactic Cluster
Survey to select genuine cluster members in a young open cluster
from the photometry in 5 passbands as well as the proper motion
by using 2MASS as first epoch. This result is extremely
promising for the remaining open clusters and star--forming
regions targeted by the GCS, including the Pleiades and
$\alpha$ Per. The second epoch conducted in the $K$ band
within the framework of the GCS will provide complete
samples of very low--mass stars and brown dwarfs over the
entire cluster with photometric and proper
motion membership. 

%
%
\section*{Acknowledgments}

NL and TRK are postdoctoral research associates funded by the UK PPARC.
We are grateful to Isabelle Baraffe and France Allard for providing 
us with the NextGen, DUSTY and COND models for the WFCAM filters.
We acknowledge Matthew Bate and Pavel Kroupa for discussion
on the mass function.
We thank our colleagues at the UK Astronomy Technology Centre,
the Joint Astronomy Centre in Hawaii, the Cambridge Astronomical Survey
and Edinburgh Wide Field Astronomy Units for building and operating
WFCAM and its associated data flow system.
The authors wish to extend special thanks to those of Hawaiian
ancestry on whose sacred mountain we are privileged to be guests.
This research has made use of the Simbad database, operated at
the Centre de Donn\'ees Astronomiques de Strasbourg (CDS), and
of NASA's Astrophysics Data System Bibliographic Services (ADS).
This publication has also made use of data products from the
Two Micron All Sky Survey, which is a joint project of the
University of Massachusetts and the Infrared Processing and
Analysis Center/California Institute of Technology, funded by
the National Aeronautics and Space Administration and the National
Science Foundation.

%
%
\bibliographystyle{mn2e}
\bibliography{../../AA/mnemonic,../../AA/biblio_old}

%
%
\appendix

\section{SQL query submitted to the WFCAM Science Archive}
\label{sql}

Initial sample selection for this Upper Sco study was made by accessing
the UKIDSS Science Verification database (UKIDSSR2) held at the WFCAM 
Science Archive. The SQL query given in the Appendix was used
to create Fig. \ref{fig_USco:ZJZcmd}.

\begin{figure*}
\begin{verbatim}
SELECT 
/*    Attribute selection:                                               */
      g.ra, g.dec, zaperMag3-yaperMag3 AS zmy, zaperMag3-japerMag3 AS zmj,
      zaperMag3-k_1aperMag3 AS zmk, yaperMag3-japerMag3 AS ymj, 
      japerMag3-haperMag3 AS jmh, haperMag3-k_1aperMag3 AS hmk,
      japerMag3-k_1aperMag3 AS jmk, zaperMag3, yaperMag3, 
      japerMag3, haperMag3, k_1aperMag3, zaperMag3Err, yaperMag3Err,
      japerMag3Err, haperMag3Err, k_1aperMag3Err

FROM 
/*    Table(s) from which to select the attributes:                      */
      gcsSource as g 
WHERE 
/*    Sample selection predicates:                             
      only Upper Sco data (no other GCS target is south of the equator)  */
      g.dec < 0.0
/*    Bright saturation cut-offs                                         */
      AND (zaperMag3 < -0.9e9 OR zaperMag3 > 11.4)
      AND yaperMag3 > 11.3 
      AND japerMag3 > 10.5 
      AND haperMag3 > 10.2 
      AND k_1aperMag3 > 9.7
/*    Limit merged passband selection to +/- 1 arcsec                    */
      AND (zXi BETWEEN -1.0 AND +1.0 OR zXi < -0.9e9)
      AND yXi BETWEEN -1.0 AND +1.0
      AND jXi BETWEEN -1.0 AND +1.0
      AND hXi BETWEEN -1.0 AND +1.0
      AND k_1Xi BETWEEN -1.0 AND +1.0
      AND (zEta BETWEEN -1.0 AND +1.0 OR zEta < -0.9e9)
      AND yEta BETWEEN -1.0 AND +1.0
      AND jEta BETWEEN -1.0 AND +1.0
      AND hEta BETWEEN -1.0 AND +1.0
      AND k_1Eta BETWEEN -1.0 AND +1.0
/*    Retain only point-like sources                                     */
      AND (zClass BETWEEN -2 AND -1 OR zClass = -9999)
      AND yClass BETWEEN -2 AND -1
      AND jClass BETWEEN -2 AND -1
      AND hClass BETWEEN -2 AND -1
      AND k_1Class BETWEEN -2 AND -1
/*    Retain only the best record when duplicated in an overlap region   */
      AND (priOrSec = 0 OR priOrSec = g.frameSetID)
\end{verbatim}
In order to get proper motion information, substitute the following for the
\verb+FROM+ clause above:
\begin{verbatim}
      gcsMergeLog AS l, Multiframe AS mj, (
         SELECT t.ra AS ra, t.dec AS dec, x.slaveObjID AS slaveObjID,
                x.masterObjID as masterObjID, t.j_m, t.h_m, t.k_m, t.jdate
         FROM   gcsSourceXtwomass_psc as x, TWOMASS..twomass_psc as t
         WHERE  x.slaveObjID=t.pts_key AND distanceMins IN (
            SELECT MIN(distanceMins) FROM gcsSourceXtwomass_psc WHERE  masterObjID=x.masterObjID
         )
      ) AS T2 RIGHT OUTER JOIN gcsSource AS g ON (g.sourceID=T2.masterObjID)
\end{verbatim}
along with the following additional \verb+WHERE+ predicates:
\begin{verbatim}
      AND g.frameSetID=l.frameSetID
      AND l.jmfID=mj.multiframeID
\end{verbatim}
and the following two additional \verb+SELECT+ions to obtain relative proper
motions with respect to 2MASS in units of milliarcsec~yr$^{-1}$:
\begin{verbatim}
      3.6e6*COS(RADIANS(g.dec))*(g.ra-T2.ra)/((mj.mjdObs - T2.jdate+2400000.5)/365.25) AS pmRA, 
      3.6e6*(g.dec-T2.dec)/((mj.mjdObs - T2.jdate+2400000.5)/365.25) AS pmDEC
\end{verbatim}
\caption[]{Structured Query Language (SQL) query used on the WFCAM Science
Archive database UKIDSSR2 to select the GCS Upper Sco sample discussed in
the paper. The query returns 133,476 rows of data. }
\label{sqlquery}
\end{figure*}

Comments in the script indicate the various selection cuts made.
Note that in the \verb+SELECT+ statement we explicitly compute colours
based on aperture magnitudes optimal for point sources. In all database
releases after science verification this is not necessary as these
colours are held by default in the merged source tables as \verb+zmyPnt+
etc. (the science verification data had colours based on the larger
Hall radius aperture magnitudes). Also, the sample selection requires
that candidates be detected in YJHK bands, but to push the sample
selection fainter and redder we are careful to allow default values
of the $Z$ photometric attributes to retain stars not detected in that
passband. Furthermore, we employ morphological classification codes
--1 (stellar) and --2 (possibly stellar) to maximise completeness.
Finally, the \verb+WHERE+ predicate 
\verb+(priOrSec=0 OR priOrSec=frameSetID)+ purges secondary duplicates
in overlap regions. For more details concerning the use of SQL and
data in the WFCAM Science Archive, see Hambly et al.\ (2006) and references
therein.

%
%
\section{Table of photometric non-members}
 \begin{table*}
 \centering
  \caption{Coordinates (J2000), $ZYJHK$ photometry, and proper
motions of 18 photometric non-members (open squares in the
colour--magnitude and colour--colour diagrams).
}
 \label{tab_USco:photNM}
 \begin{tabular}{c c c c c c c c c}
 \hline
R.A.\ & Dec.\  &  $Z$  &  $Y$  &  $J$  &  $H$  & $K$ & $\mu_{\alpha}cos\delta$ & $\mu_{\delta}$ \\ \hline
16 06 30.92 & $-$22 58 15.2 & 16.658$\pm$0.011 & 15.990$\pm$0.008 & 14.985$\pm$0.004 & 13.628$\pm$0.003 & 12.414$\pm$0.002 &   99.9 &   99.9 \\
16 07 03.05 & $-$23 31 46.2 & 12.024$\pm$0.001 & 11.539$\pm$0.001 & 10.982$\pm$0.001 & 10.393$\pm$0.000 &  9.792$\pm$0.000 &   10.2 &  $-$24.9 \\
16 07 25.70 & $-$23 14 58.0 & 20.893$\pm$0.398 & 19.434$\pm$0.138 & 18.790$\pm$0.128 & 17.960$\pm$0.111 & 17.816$\pm$0.178 &   99.9 &   99.9 \\
16 07 27.57 & $-$21 54 42.7 & 17.010$\pm$0.015 & 16.198$\pm$0.010 & 15.588$\pm$0.007 & 14.990$\pm$0.006 & 14.564$\pm$0.009 &   99.9 &   99.9 \\
16 07 29.54 & $-$23 08 22.3 & 12.815$\pm$0.001 & 12.365$\pm$0.001 & 11.749$\pm$0.001 & 10.970$\pm$0.000 & 10.188$\pm$0.000 &   99.9 &   99.9 \\
16 08 20.24 & $-$23 33 41.7 & 20.910$\pm$0.387 & 19.780$\pm$0.198 & 18.689$\pm$0.118 & 18.291$\pm$0.148 & 18.387$\pm$0.283 &   99.9 &   99.9 \\
16 08 39.70 & $-$23 06 57.8 & 18.487$\pm$0.042 & 15.984$\pm$0.008 & 15.596$\pm$0.007 & 15.040$\pm$0.007 & 14.889$\pm$0.011 &   99.9 &   99.9 \\
16 10 51.82 & $-$23 33 26.3 & 16.359$\pm$0.009 & 15.581$\pm$0.006 & 14.969$\pm$0.005 & 14.343$\pm$0.005 & 13.985$\pm$0.006 &  $-$67.9 & $-$110.7 \\
16 11 08.90 & $-$23 08 28.3 & 11.664$\pm$0.001 & 11.361$\pm$0.001 & 10.783$\pm$0.000 & 10.499$\pm$0.000 & 10.051$\pm$0.000 &  $-$16.6 &  $-$32.0 \\
16 11 24.88 & $-$22 52 53.3 & 17.715$\pm$0.023 & 16.735$\pm$0.016 & 16.007$\pm$0.011 & 15.315$\pm$0.008 & 14.875$\pm$0.011 &   99.9 &   99.9 \\
16 12 05.64 & $-$22 24 12.4 & 19.915$\pm$0.157 & 18.568$\pm$0.063 & 17.546$\pm$0.038 & 16.990$\pm$0.033 & 16.500$\pm$0.045 &   99.9 &   99.9 \\
16 12 34.14 & $-$21 44 50.2 & 13.461$\pm$0.002 & 12.831$\pm$0.002 & 12.387$\pm$0.001 & 11.704$\pm$0.001 & 11.205$\pm$0.001 &   $-$6.0 &  $-$26.7 \\
16 13 01.40 & $-$21 42 54.5 & 18.135$\pm$0.035 & 17.201$\pm$0.020 & 16.325$\pm$0.011 & 15.613$\pm$0.010 & 15.065$\pm$0.013 &   99.9 &   99.9 \\    
16 13 39.04 & $-$22 33 14.4 & 19.558$\pm$0.109 & 18.399$\pm$0.061 & 17.418$\pm$0.042 & 16.836$\pm$0.033 & 16.227$\pm$0.039 &   99.9 &   99.9 \\
16 14 50.26 & $-$23 32 39.9 & 12.758$\pm$0.001 & 12.218$\pm$0.001 & 11.605$\pm$0.001 & 10.803$\pm$0.000 & 10.067$\pm$0.000 &  $-$18.2 &  $-$16.4 \\
16 15 03.04 & $-$23 52 50.3 & 17.473$\pm$0.019 & 16.650$\pm$0.013 & 15.932$\pm$0.009 & 15.286$\pm$0.010 & 14.842$\pm$0.012 &    99.9  &  99.9    \\
16 15 27.63 & $-$23 52 34.2 & 14.833$\pm$0.004 & 13.989$\pm$0.003 & 13.271$\pm$0.002 & 12.442$\pm$0.001 & 11.808$\pm$0.001 &    0.1 &  $-$27.6 \\
16 16 22.51 & $-$23 38 42.6 & 20.331$\pm$0.220 & 18.841$\pm$0.080 & 17.991$\pm$0.058 & 17.247$\pm$0.060 & 16.989$\pm$0.083 &   99.9 &   99.9 \\
 \hline
 \end{tabular}
\end{table*}

%
%
\section{Table of proper motion non-members}
 \begin{table*}
 \centering
  \caption{Coordinates (J2000), $ZYJHK$ photometry, and proper
motions of 26 proper motion non-members (open diamonds in the
colour-magnitude and colour-colour diagrams). 
}
 \label{tab_USco:PM_NM}
 \begin{tabular}{c c c c c c c c c}
 \hline
R.A.\ & Dec.\  &  $Z$  &  $Y$  &  $J$  &  $H$  & $K$ & $\mu_{\alpha}cos\delta$ & $\mu_{\delta}$ \\ \hline
16 06 57.53 & $-$23 03 27.1 & 11.759$\pm$0.001 & 11.404$\pm$0.001 & 10.901$\pm$0.000 & 10.341$\pm$0.000 &  9.970$\pm$0.000 &  $-$11.8 &    9.7 \\
16 07 03.05 & $-$23 31 46.2 & 12.024$\pm$0.001 & 11.539$\pm$0.001 & 10.982$\pm$0.001 & 10.393$\pm$0.000 &  9.792$\pm$0.000 &   10.2 &  $-$24.9 \\
16 07 27.57 & $-$21 54 42.7 & 17.010$\pm$0.015 & 16.198$\pm$0.010 & 15.588$\pm$0.007 & 14.990$\pm$0.006 & 14.564$\pm$0.009 &  $-$83.1 &  $-$51.3 \\
16 07 48.27 & $-$21 39 03.9 & 13.962$\pm$0.002 & 13.300$\pm$0.002 & 12.686$\pm$0.001 & 12.118$\pm$0.001 & 11.778$\pm$0.001 &   $-$2.8 &   $-$4.5 \\
16 08 27.33 & $-$22 17 29.3 & 12.756$\pm$0.001 & 12.183$\pm$0.001 & 11.630$\pm$0.001 & 10.991$\pm$0.000 & 10.586$\pm$0.001 &    0.4 &   $-$8.1 \\
16 08 28.69 & $-$21 37 20.0 & 12.758$\pm$0.001 & 12.154$\pm$0.001 & 11.598$\pm$0.001 & 11.073$\pm$0.000 & 10.685$\pm$0.001 &  $-$31.4 &  $-$20.5 \\
16 08 34.55 & $-$22 11 55.9 & 13.772$\pm$0.002 & 13.067$\pm$0.002 & 12.500$\pm$0.001 & 11.943$\pm$0.001 & 11.556$\pm$0.001 &   $-$7.1 &   $-$5.2 \\
16 09 05.67 & $-$22 45 16.7 & 17.341$\pm$0.019 & 16.049$\pm$0.008 & 15.212$\pm$0.006 & 14.558$\pm$0.006 & 13.987$\pm$0.007 &  $-$25.7 &   $-$9.9 \\
16 09 07.75 & $-$23 39 54.5 & 13.438$\pm$0.002 & 12.629$\pm$0.001 & 12.032$\pm$0.001 & 11.476$\pm$0.001 & 11.107$\pm$0.001 &  $-$17.1 &   $-$3.7 \\
16 09 08.83 & $-$22 17 46.9 & 14.124$\pm$0.003 & 13.466$\pm$0.002 & 12.903$\pm$0.001 & 12.403$\pm$0.001 & 12.046$\pm$0.001 &  $-$36.5 &  $-$29.6 \\
16 09 32.29 & $-$22 29 36.0 & 12.421$\pm$0.001 & 11.930$\pm$0.001 & 11.383$\pm$0.001 & 10.764$\pm$0.000 & 10.445$\pm$0.001 &   $-$6.1 &    8.2 \\
16 09 37.85 & $-$21 23 19.0 & 12.638$\pm$0.001 & 12.170$\pm$0.001 & 11.700$\pm$0.001 & 11.187$\pm$0.001 & 10.877$\pm$0.001 &   $-$3.6 & $-$241.2 \\
16 10 00.18 & $-$23 12 19.3 & 15.947$\pm$0.007 & 15.204$\pm$0.005 & 14.530$\pm$0.004 & 13.977$\pm$0.004 & 13.545$\pm$0.004 &  $-$33.5 &  $-$14.2 \\
16 10 02.67 & $-$23 44 39.5 & 12.413$\pm$0.001 & 11.758$\pm$0.001 & 11.204$\pm$0.001 & 10.682$\pm$0.000 & 10.265$\pm$0.001 & $-$177.8 &   89.7 \\
16 10 02.86 & $-$23 44 40.9 & 11.900$\pm$0.001 & 11.395$\pm$0.001 & 10.862$\pm$0.000 & 10.427$\pm$0.000 &  9.949$\pm$0.000 &  252.5 & $-$145.7 \\
16 10 19.43 & $-$23 31 08.9 & 14.842$\pm$0.004 & 14.123$\pm$0.003 & 13.532$\pm$0.002 & 12.967$\pm$0.002 & 12.571$\pm$0.002 &   $-$5.1 &   $-$2.5 \\
16 10 51.82 & $-$23 33 26.3 & 16.359$\pm$0.009 & 15.581$\pm$0.006 & 14.969$\pm$0.005 & 14.343$\pm$0.005 & 13.985$\pm$0.006 &  $-$67.9 & $-$110.7 \\
16 14 01.73 & $-$22 58 48.7 & 13.654$\pm$0.002 & 12.981$\pm$0.002 & 12.422$\pm$0.001 & 11.882$\pm$0.001 & 11.515$\pm$0.001 &  312.7 & $-$410.9 \\
16 14 12.36 & $-$22 19 13.2 & 12.469$\pm$0.001 & 11.982$\pm$0.001 & 11.446$\pm$0.001 & 11.057$\pm$0.000 & 10.568$\pm$0.001 &    8.8 &    0.2 \\
16 14 25.25 & $-$23 50 15.1 & 13.520$\pm$0.002 & 12.875$\pm$0.001 & 12.341$\pm$0.001 & 11.804$\pm$0.001 & 11.422$\pm$0.001 &  $-$14.2 &   $-$5.0 \\
16 14 46.84 & $-$23 41 57.2 & 16.314$\pm$0.009 & 15.494$\pm$0.006 & 14.835$\pm$0.004 & 14.251$\pm$0.004 & 13.815$\pm$0.005 &  $-$34.6 &  $-$14.9 \\
16 15 16.02 & $-$23 45 10.4 & 13.587$\pm$0.002 & 12.844$\pm$0.001 & 12.230$\pm$0.001 & 11.656$\pm$0.001 & 11.272$\pm$0.001 &  $-$11.0 &    3.9 \\
16 15 16.66 & $-$23 40 46.3 & 17.968$\pm$0.030 & 16.552$\pm$0.013 & 15.621$\pm$0.008 & 14.911$\pm$0.007 & 14.269$\pm$0.008 &  $-$35.1 &  $-$11.1 \\
16 15 38.43 & $-$23 41 56.0 & 13.399$\pm$0.002 & 12.760$\pm$0.001 & 12.208$\pm$0.001 & 11.584$\pm$0.001 & 11.267$\pm$0.001 &  $-$32.1 &  $-$16.2 \\
16 16 20.11 & $-$23 44 14.3 & 12.083$\pm$0.001 & 11.504$\pm$0.001 & 10.976$\pm$0.001 & 10.533$\pm$0.000 & 10.144$\pm$0.000 &  $-$33.2 &  $-$11.3 \\
16 16 26.20 & $-$23 50 48.8 & 12.051$\pm$0.001 & 11.601$\pm$0.001 & 11.018$\pm$0.001 & 10.474$\pm$0.000 &  9.968$\pm$0.000 &  $-$10.0 &    1.1 \\
 \hline
 \end{tabular}
\end{table*}



\label{lastpage}

\end{document}